\pdfoutput=1

\documentclass[11pt,twoside,a4paper,cmspaper,final,collab]{cms-tdr}
\def\svnVersion{179510}\def\cmsCernNo{2012-316}\def\cmsCernDate{\today}\def\cmsMessage{Submitted to Physical Review D}

\begin{document}\cmsNoteHeader{SUS-11-028}

\hyphenation{had-ron-i-za-tion}
\hyphenation{cal-or-i-me-ter}
\hyphenation{de-vices}

\RCS$Revision: 165121 $
\RCS$HeadURL: svn+ssh://svn.cern.ch/reps/tdr2/papers/SUS-11-028/trunk/SUS-11-028.tex $
\RCS$Id: SUS-11-028.tex 165121 2013-01-16 15:36:34Z alverson $
\newlength\cmsFigWidth
\ifthenelse{\boolean{cms@external}}{\setlength\cmsFigWidth{0.85\columnwidth}}{\setlength\cmsFigWidth{0.6\textwidth}}
\ifthenelse{\boolean{cms@external}}{\providecommand{\cmsLeft}{top}}{\providecommand{\cmsLeft}{left}}
\ifthenelse{\boolean{cms@external}}{\providecommand{\cmsRight}{bottom}}{\providecommand{\cmsRight}{right}}
\ifthenelse{\boolean{cms@external}}{\providecommand{\cmsThreeLeft}{top}}{\providecommand{\cmsThreeLeft}{top}}
\ifthenelse{\boolean{cms@external}}{\providecommand{\cmsThreeCenter}{center}}{\providecommand{\cmsThreeCenter}{bottom left}}
\ifthenelse{\boolean{cms@external}}{\providecommand{\cmsThreeRight}{bottom}}{\providecommand{\cmsThreeRight}{bottom right}}
\newcommand{\ETslashSection}{\texorpdfstring{\ETslash}{Missing ET}\xspace}
\newcommand{\ETslashSubSection}{\ETslashSection\xspace}
\newcommand{\CLs}{\ensuremath{\text{CL}_\text{s}}\xspace}

\newcommand\mht {\not\!\!H_\mathrm{T}}
\newcommand\mpt {\not\!\!P_\mathrm{T}}
\newcommand\ptrel{p_T^\text{rel}}
\newcommand\Phimpt{\phi(\mpt)}
\newcommand\ttbarEleNu{\ensuremath{\ttbar\to\Pe \nu +\text{jets}}}
\newcommand\ttbarMuNu{\ensuremath{\ttbar\to\mu \nu +\text{jets}}}
\newcommand\ttbarTAUHNu{\ensuremath{\ttbar\to \tau_\mathrm{h} \nu +\text{jets}}}
\newcommand\ttbarTAUMuNu{\ensuremath{\ttbar\to  \tau_{\mu} \nu +\text{jets}}}
\newcommand\ttbarMUTAUHNu{\ensuremath{\ttbar\to\mu \nu + \tau_\mathrm{h} \nu +\text{jets}}}
\newcommand\ttbarDITAUHNu{\ensuremath{\ttbar\to \tau_\mathrm{h}+ \tau_\mathrm{h} \nu +\text{jets}}}
\newcommand\znunu{\ensuremath{\cPZ \rightarrow \nu \Pagn} }
\newcommand\zmumu{\ensuremath{\cPZ \rightarrow \mu \mu} }
\newcommand\ztautau{\ensuremath{\cPZ\rightarrow \tau \tau} }
\newcommand\zee{\ensuremath{\cPZ\rightarrow \Pe \Pe} }
\newcommand\zll{\ensuremath{\cPZ\rightarrow \ell \ell} }
\newcommand\wtaunu{\ensuremath{\PW\rightarrow \tau \nu} }
\newcommand\wmunu{\ensuremath{\PW\rightarrow \mu \nu} }
\newcommand\wtauhad{\ensuremath{\PW\rightarrow \tau_\mathrm{h} \nu} }
\newcommand\wtaumu{\ensuremath{\PW\rightarrow \tau_{\mu} \nu} }
\newcommand\defMPT{\ensuremath{ \mathrm{MPT} = \mpt = \abs{ - \sum_\text{tracks} \vec{p_\mathrm{T}}}}}
\newcommand\defHT{\ensuremath{ H_\mathrm{T} = \abs{ - \sum_{i} \vec{jet_{i}}}}}
\newcommand\ppj{photon+jets }
\newcommand\mpj{MET+jets }
\newcommand\wpj{W+jets }
\newcommand\zpj{Z+jets }
\newcommand\fm{FILLME}
\newcommand\G{\ensuremath{\textmd{GeV}}}

\newcommand\TTJets{\ensuremath{\textmd{T}\bar{\textmd{T}}+\textmd{Jets}}}
\newcommand\WJets{\ensuremath{\textmd{W}+\textmd{Jets}}}
\newcommand\DYJets{\ensuremath{\textmd{DY}+\textmd{Jets}}}

\newcommand{\zeroTag}{$0$ b-tag\xspace}
\newcommand{\oneTag}{$1$ b-tag\xspace}
\newcommand{\geqOneTag}{${\ge}1$ b-tag\xspace}
\newcommand{\twoTag}{${\ge}2$ b-tag\xspace}
\newcommand{\yMET}{\ensuremath{Y_{\rm MET}}\xspace}
\newcommand{\mt}{\ensuremath{m_{\rm T}}\xspace}
\newcommand{\mthree}{\ensuremath{M_{3}}\xspace}

\newcommand{\Wjets}{\PW+jets\xspace}
\newcommand{\Wpjets}{\PWp+jets\xspace}
\newcommand{\Wmjets}{\PWm+jets\xspace}

\newcommand{\HTmiss}{\ensuremath{H_{\mathrm{T}}^{\text{miss}}}\xspace}

\newcommand{\HTslash}{\ensuremath{\slash\mkern-12mu{H}_{\text{T}}}\xspace}

\newcommand{\pthat}{\mbox{$\hat{p}_\mathrm{T}$}\xspace}
\ifthenelse{\boolean{cms@external}}{%
\newcommand{\scotchrule[1]}{\centering\begin{ruledtabular}\begin{tabular}{#1}}
\newcommand{\donescotchrule}{\end{tabular}\end{ruledtabular}}
}{
\newcommand{\scotchrule[1]}{\centering\begin{tabular}{#1}\hline}
\newcommand{\donescotchrule}{\hline\end{tabular}}
}

\cmsNoteHeader{SUS-11-028} % This is over-written in the CMS environment: useful as preprint no. for export versions
\title{Search for supersymmetry in final states with a single lepton, b-quark jets, and missing transverse energy in proton-proton collisions at \texorpdfstring{$\sqrt{s}=7\TeV$}{sqrt(s) = 7 TeV}}

\date{\today}

\abstract{A search motivated by supersymmetric models with light top squarks is presented using proton-proton collision data recorded with the CMS detector at a center-of-mass energy of $\sqrt{s}=7$\TeV during 2011, corresponding to an integrated luminosity of 4.98\fbinv. The analysis is based on final states with a single lepton, b-quark jets, and missing transverse energy.
Standard model yields are predicted from data using two different approaches. The observed event numbers are
found to be compatible with these predictions. Results are interpreted in the context of the constrained minimal supersymmetric standard model and of a simplified model with four top quarks in the final state.
}

\hypersetup{%
pdfauthor={CMS Collaboration},%
pdftitle={Search for supersymmetry in final states with a single lepton, b-quark jets, and missing transverse energy in proton-proton collisions at sqrt(s) = 7 TeV},%
pdfsubject={CMS},%
pdfkeywords={CMS, physics, SUSY}}

\maketitle %maketitle comes after all the front information has been supplied

\section{Introduction}
In this paper we describe a search for supersymmetry (SUSY) in final states with a single electron or muon,
multiple jets, including some identified as originating from \cPqb\ quarks (\cPqb\ jets), and missing transverse energy.
The search is based on the full set of data recorded with the Compact Muon Solenoid (CMS) experiment in
proton-proton collisions at a center-of-mass energy of $\sqrt{s} = 7$\TeV during 2011, which
corresponds to an integrated luminosity of $4.98 \pm 0.11$\fbinv.

The search for new physics phenomena in events with third-generation quarks at the Large Hadron
Collider (LHC) is motivated by various extensions~\cite{Martin:1997ns, Flacco:2010rg,phenomLittleHiggs,ExtraDimensions,compositeHiggs} of the standard model (SM). Among these,
supersymmetric models are
regarded as attractive, because they can resolve the hierarchy problem and may permit the unification of
the electroweak and strong interactions ~\cite{ref:SUSY0,ref:SUSY1,ref:SUSY2,ref:SUSY3,ref:SUSY4}.

Supersymmetry predicts that for each particle in the SM there exists a partner particle, often
referred to as a sparticle, with identical gauge quantum numbers but with a spin that differs by $1/2$.
Assuming $R$~parity conservation~\cite{FarrarEtAl}, sparticles are produced in pairs, and
their decay chains terminate with the lightest supersymmetric particle (LSP).
In some scenarios the LSP is the lightest neutralino ($\PSGczDo$), a heavy, electrically
neutral, weakly interacting particle, which is a viable dark-matter candidate. In these scenarios,
SUSY events are characterized by missing transverse energy in the final state.

In several SUSY scenarios, particularly motivated by naturalness of the spectrum~\cite{Papucci:2011wy,Brust:2011tb},
top (\sTop) or bottom (\sBot) squarks may be copiously produced at the LHC.
This may happen by direct squark production, e.g.,
$\Pp\Pp \to \sTop\,\sTop^{*}\,\to\,\cPqt\,\cPaqt\,\PSGczDo\,\PSGczDo$.
If the mass of the gluino ($\PSg$) is larger than the masses
of the third-generation squarks, but lighter than the squarks of the first two generations,  the gluino may dominantly decay into the third-generation squarks, e.g.,
        $\PSg\,\rightarrow\,\cPqt\,\sTop^*\,\rightarrow\,\cPqt\,\cPaqt\,\PSGczDo$.
Hence gluino pair production can lead to events containing four third-generation quarks,
resulting in an excess of events with large \cPqb-jet multiplicities, which is exploited by dedicated analyses \cite{Chatrchyan:2012sa, Chatrchyan:2012jx, Chatrchyan:1474300, Aad:2012ar, Aad:2012si, Aad:2012pq, ATLAS:2012ah}.

The decay chains of the strongly interacting particles predicted by these models result in a high level of hadronic activity, characterized by a large number of high-energy jets.
In addition, isolated leptons may originate from leptonically decaying top quarks and two- or three-body decays of neutralinos and charginos.

The search is performed in signal regions defined using the scalar sum of the jet transverse
momenta \HT, the missing transverse energy \ETslash, and the \cPqb-jet multiplicity.
The dominant SM background processes contributing to the search topology are top-quark pair (\ttbar) production
and inclusive W-boson production in association with energetic jets (\Wjets).
Smaller  contributions are due to single-top production, QCD multijet events (QCD), and Drell-Yan (DY) production and decay to lepton pairs in which one lepton goes undetected. While simulation provides a good description
of these contributions, more reliable estimates of the backgrounds can be obtained from data.

To evaluate the SM background, two complementary data-based approaches are used.
In the first approach, templates for the \ETslash spectra in \Wmjets, \Wpjets, and \ttbar production are extracted from
the inclusive single-lepton sample by a simultaneous fit to the 0, 1, and ${\ge}2$ \cPqb-jet subsamples.
This fit involves the convolution of a model for the true \ETslash distribution with detector effects
determined using data in control regions at low \HT. Predictions in several signal regions defined by different
selections on \HT, \ETslash, and for 0, 1, and ${\ge}2$ identified \cPqb\ jets are obtained by applying the
templates at high values of \HT after normalization in background-dominated regions at low \ETslash.
The second approach, a factorization method, predicts the expected number of background events in a subsample with high \HT and \yMET, where $Y_\mathrm{MET} = \ETslash / \sqrt{\HT}$ is an approximate measure of the \ETslash significance.
Since \HT and \yMET are only weakly correlated, the estimate can be obtained using a factorization approach based on three background-dominated control regions and can be calculated independently for
different \cPqb-jet multiplicities. Therefore, it naturally provides an estimate for
a selection with ${\ge}3$ identified \cPqb-jets, yielding a better signal-to-noise ratio for SUSY models with
many (at least 3) \cPqb\ jets. The use of a background estimation technique based on data reduces the uncertainty on the prediction by more than a factor of two.
While both methods use the \HT and \ETslash variables, they have only a small overlap in their control and signal
regions, both in the SM and in the signal scenarios, and are therefore complementary.

The analyses presented here are not limited to a particular theory.
However, the constrained minimal supersymmetric extension of the standard model (CMSSM)~\cite{PhysRevLett.49.970, PhysRevD.27.2359}
is chosen as a benchmark to illustrate the sensitivity of this search for new-physics processes.
The template method in the 0, 1, and ${\ge}2$ \cPqb-jet subsamples shows the best sensitivity in the parameter plane of this model.
A scenario involving four top quarks in the final state is used as the second benchmark.
It is implemented as a scenario in the context of simplified model spectra (SMS)~\cite{ArkaniHamed:2007fw,Alves:2011wf,Alwall:2008ag}.
The factorization method with the ${\ge}3$ \cPqb-jet subsample is best suited for this topology.

A brief description of the CMS detector is given in Section~\ref{sec:CMSDetector}.
The datasets and simulated event samples used in this search are discussed in Section~\ref{sec:EventSamples}.
In Section~\ref{sec:EventSelection} the preselection of physics objects and events is outlined.
The \ETslash template and factorization methods are described in Sections~\ref{sec:SR_BG_METT} and \ref{sec:SR_BG_ABCD}, respectively.
Results are presented in Section~\ref{sec:Result} and interpreted in Section~\ref{sec:Interpretation}.
Finally a summary is given in Section~\ref{sec:Conclusion}.

\section{The CMS Detector}
\label{sec:CMSDetector}
The CMS detector is a multipurpose
apparatus designed to allow the study of high transverse momentum (\pt) processes in proton-proton collisions, as well as a broad range of phenomena in heavy-ion collisions.
The CMS coordinate system is defined with the origin at the center of the detector and the $z$ axis along the counterclockwise beam direction, with $\phi$ the azimuthal angle (measured in radians), $\theta$ the polar angle, and $\eta=-\ln[\tan(\theta/2)]$ the pseudorapidity.

The central feature of the detector is a superconducting solenoid,
13~m in length and 6~m in diameter, which provides an axial magnetic field of 3.8\unit{T}.
Within the magnet are the silicon pixel and strip detectors for charged-particle tracking, a lead-tungstate crystal electromagnetic calorimeter for measurements of photons, electrons, and the electromagnetic component
of jets, and a hadron calorimeter, constructed from scintillating tiles and brass absorbers, for jet energy measurements.
The tracker covers the region $|\eta| < 2.5$ and the calorimeters $|\eta|<3.0$.
A quartz-steel forward calorimeter using Cherenkov radiation extends
the coverage to $|\eta| \le 5$. The detector is nearly hermetic, allowing for energy-balance measurements
in the plane transverse to the beam direction.
Outside the magnet is the muon system, comprising drift-tube, cathode-strip,
and resistive-plate detectors, all interleaved with steel absorbers acting as a magnetic flux return.
A detailed description of the CMS detector can be found elsewhere~\cite{ref:CMS}.

\section{Event Samples}
\label{sec:EventSamples}

The events are selected with triggers requiring
the presence of a muon or electron with large transverse momentum $\pt$ in
association with significant hadronic activity, quantified by $\HT^\text{trigger}$, the value of \HT calculated at the trigger level.
In the second part of the year a requirement on $\HTslash^\text{trigger}$, the magnitude of the vectorial sum of the transverse momenta of jets, was added.

In order not to exceed the maximum possible rate of data acquisition and processing, trigger thresholds were
raised with increasing LHC luminosity, resulting in a threshold for the muon transverse momentum \pt from 8\GeV to 15\GeV, and for electrons from 10\GeV to 15\GeV.
The requirement on the hadronic activity was raised from $\HT^\text{trigger} > 200\GeV$
to $\HT^\text{trigger} > 300$\GeV in the muon and to $\HT^\text{trigger} > 250\GeV$ in the electron channel.
The requirement on $\HTslash^\text{trigger}$ was introduced with a threshold of 20\GeV that was later raised to 40\GeV.

Simulated event samples are produced using different event generators and the \GEANTfour\ package~\cite{GEANT4} for detector
simulation, except for the scans of CMSSM and SMS parameter space discussed below.
The production and decay of \ttbar pairs or vector bosons in
association with energetic jets are generated using the \MADGRAPH~5.1.1~\cite{Alwall:2007st} generator.
The produced parton events are then passed to the \PYTHIA~6.4.24~\cite{Sjostrand:2006za} program
with tune Z2~\cite{ref:TuneZ2} for simulating parton showers, multiple interactions,
and fragmentation processes. The decay of $\tau$ leptons is simulated using the \TAUOLA~27.121.5~\cite{Was:2000st} program. The production and decay
of single top quarks and antiquarks are simulated with the \POWHEG~301~\cite{Alioli:2009je,Re:2010bp}
and \TAUOLA generators interfaced to \PYTHIA. Multijet QCD production is simulated with \PYTHIA.

Mass spectra and branching fractions of SUSY particles are calculated at the electroweak scale using the renormalization equations
implemented in the \textsc{SoftSusy} package~\cite{Allanach:2001kg}, interfaced to \PYTHIA.
Two low-mass scenarios~\cite{PTDR2} are used as CMSSM benchmark points to illustrate possible yields:
the first one is referred to as LM6 ($m_0 = 85\GeV$, $m_{1/2} = 400\GeV$, $A_0 = 0\GeV$, $\tan\beta = 10$, $\mu>0$), and the second one as LM8 ($m_0 = 500\GeV$, $m_{1/2} = 300\GeV$, $A_0 = -300\GeV$, $\tan\beta = 10$, $\mu>0$). In other event topologies these points have been experimentally excluded~\cite{Chatrchyan:2012sa, Chatrchyan:2012jx, Chatrchyan:2012mfa}.

A scan in the CMSSM parameter space is performed for a fixed set of parameters: $A_0$, $\tan{\beta}$,
and sign\,$\mu$, where a grid in the $m_0$\,-\,$m_{1/2}$ plane is defined by
variation of $m_0$ and $m_{1/2}$ in steps of 20\GeV. For each point, 10\,000 $\Pp\Pp$ events are generated.

\begin{figure}[htb]
 \begin{center}
    \includegraphics[width=\cmsFigWidth]{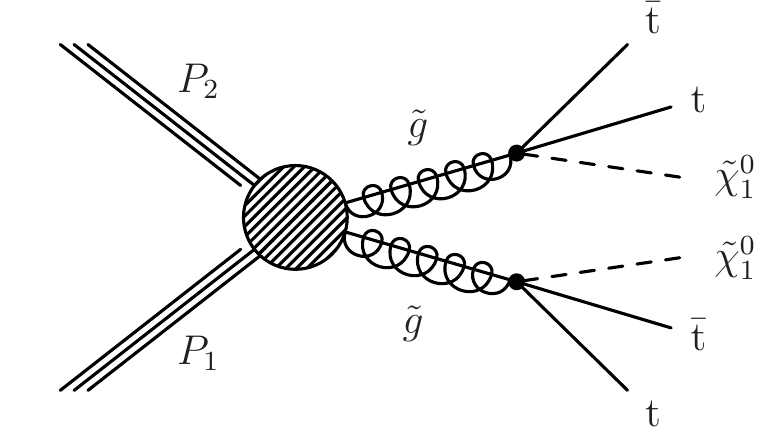}
 \end{center}
 \caption{
    Diagram for the simplified model used in this paper.
   }\label{fig:T1tttt}
\end{figure}

In addition, the results are interpreted in the context of the simplified model shown in Fig.~\ref{fig:T1tttt}.
It contains the pair production of gluinos, which subsequently decay with branching fraction $\mathcal{B}(\PSg \rightarrow \ttbar  + \PSGczDo) = 1$.
For each point on a $25\GeV \times 25\GeV$ grid in the parameter plane of the gluino and $\PSGczDo$ masses, 50\,000 events are simulated.
The events in the CMSSM and SMS scans are generated using a fast detector simulation \cite{Abdullin:1328345} rather than the \GEANTfour package.

\section{Event Selection}
\label{sec:EventSelection}

The primary vertex must satisfy a set of quality requirements, including the restriction that the longitudinal and transverse distances of the primary vertex from the nominal interaction point be less than 24\unit{cm} and 2\unit{cm}, respectively.

Muon candidates~\cite{ref:PAS-MUO-10-002} are required to have $\pt(\mu)>20$\GeV and $|\eta|<2.1$.
The reconstructed track of a muon candidate must have an impact parameter less than 0.02~cm in the transverse plane and less than 1.0\unit{cm} along the beam axis, where the impact parameter is the distance of the track trajectory to the primary vertex at the point of closest approach in the transverse plane.
To suppress background contributions from muons originating from heavy-flavor
quark decays, the muon is required to be isolated within a cone of size $\Delta R =0.3$, with $\Delta R = \sqrt{(\Delta\eta)^2+ (\Delta\phi)^2}$.
The relative combined isolation of the muon is defined as $I^\text{comb}_\text{rel}=\sum_{\Delta R<0.3}(\et + \pt)/\pt(\mu)\,$,
where the sum is over the transverse energy \et (as measured in the electromagnetic and hadron calorimeters) and the transverse
momentum \pt (as measured in the silicon tracker) of all reconstructed objects within this cone, excluding the track itself.
Muons are required to satisfy $I^\text{comb}_\text{rel}<0.1$.

Electron candidates~\cite{ref:PAS-EGM-10-004} are restricted to $\pt>20$\GeV and $|\eta|<2.4$,
excluding the barrel-endcap transition region ($1.44 <|\eta|<1.57$). The reconstructed track of an electron candidate must fulfill the
same impact parameter requirements as the muon tracks described above, as well as a set of quality and photon-conversion rejection criteria.
The relative combined isolation variable,
similar to that defined in the muon case, must satisfy $I^\text{comb}_\text{rel}<0.07$ in the barrel region and
$I^\text{comb}_\text{rel}<0.06$ in the endcaps.

Exactly one selected muon or electron is required to be present in the event.
Events with a second lepton passing looser selection criteria are rejected.

The reconstruction of jets is based on the CMS particle-flow algorithm~\cite{ref:PAS-PFT-09-001}, which identifies and reconstructs charged hadrons, electrons, muons, photons, and neutral hadrons.
Extra energy clustered into jets due to additional, simultaneous pp collisions (``pileup'') is taken into account with an event-by-event correction
to the jet four-vectors \cite{Cacciari:2011ma}. Therefore, the pileup does not have a strong influence on this analysis.
Jets are reconstructed from particle-flow candidates using the anti-\kt clustering
algorithm \cite{ref:antikt} with distance parameter 0.5. Corrections are applied on
the raw jet energy to obtain a uniform response across the detector in $\eta$ and an
absolute calibrated response in $\pt$ \cite{Chatrchyan:2011ds}. Each event is required to contain at
least three jets with $\pt>40\GeV$ and $|\eta|< 2.4$ that are spatially separated from a selected
muon or electron by $\Delta R>0.3$ and that satisfy quality criteria in order to suppress noise and
spurious calorimeter energy deposits.

The identification of \cPqb\ jets (``\cPqb-tagging'')~\cite{ref:PAS-BTV-11-004} is performed with two complementary approaches.
In the first approach, the distance between a reconstructed secondary vertex with two or more associated tracks and
the primary interaction point, normalized to its uncertainty, is used (simple secondary-vertex algorithm).
This algorithm has been shown to be particularly robust against variation in the running conditions and is used for the template method.
In the second approach, jets are tagged as \cPqb\ jets if they have at least two tracks with an impact parameter divided by its uncertainty that is greater than 3.3 (track counting algorithm).
This algorithm is highly efficient at high jet \pt\ and is used for the factorization method.
At the chosen operating points, the efficiency to tag \cPqb\ jets is approximately 60 to 70\%, with a misidentification rate for light-quark- or gluon-initiated jets of a few percent.
The b-tagging efficiencies and mistagging rates (the efficiency of tagging a c-quark jet, light-quark jet, or gluon jet as \cPqb\ jet) have been measured up to jet \pt of 670\GeV for both methods.

The missing transverse energy $\ETslash$ is reconstructed as the magnitude of the sum of the transverse momentum vectors of all particle-flow objects with $|\eta| < 4.7$.
The quantity $H_{\mathrm{T}}$, a measure of the total hadronic activity, is calculated as the sum of the transverse momenta of all jets passing the selection.
Since SUSY models predict events with large hadronic activity and large amounts of missing energy,
the final search regions for the two methods are defined by stringent selections on $\HT$ and $\ETslash$
and by the number of identified \cPqb\ jets, as described in the following two sections.

These selection steps define a sample that matches the trigger requirements and the expected characteristics of signal events, while retaining a sufficient number of events to allow evaluation of the background.

The trigger and lepton-reconstruction efficiencies are measured from data.
The determination of the trigger efficiency is performed separately for each component of the trigger: the leptonic, the $\HT^\text{trigger}$, and the $\HTslash^\text{trigger}$ selection.
The leptonic trigger selection is found to be 97--98\% efficient after the offline requirements, for all running periods.
The  $\HT^\text{trigger}$ requirement, and the $\HTslash^\text{trigger}>20\GeV$ trigger requirement used for the first part of the running period, are both more than 99\% efficient.
The $\HTslash^\text{trigger} > 40\GeV$ requirement used for latter part of the running period is around 80\% efficient for $\ETslash$ values of 60\GeV, becoming fully efficient for $\ETslash > 80\GeV$.

The offline lepton reconstruction, identification, and isolation efficiencies are measured with a "tag-and-probe" method~\cite{PAS-EGM-07-001}, using dileptons with invariant mass close to the \cPZ~peak.
The measured efficiencies have been compared to simulation as a function of \pt, $\eta$, and the number of reconstructed primary vertices and jets in the event.
The total lepton efficiency in data is described by simulation to a relative accuracy within 3\%.
\section{The \ETslashSection Template Method}
\label{sec:SR_BG_METT}

\begin{figure}
\begin{center}
\includegraphics[width=\cmsFigWidth]{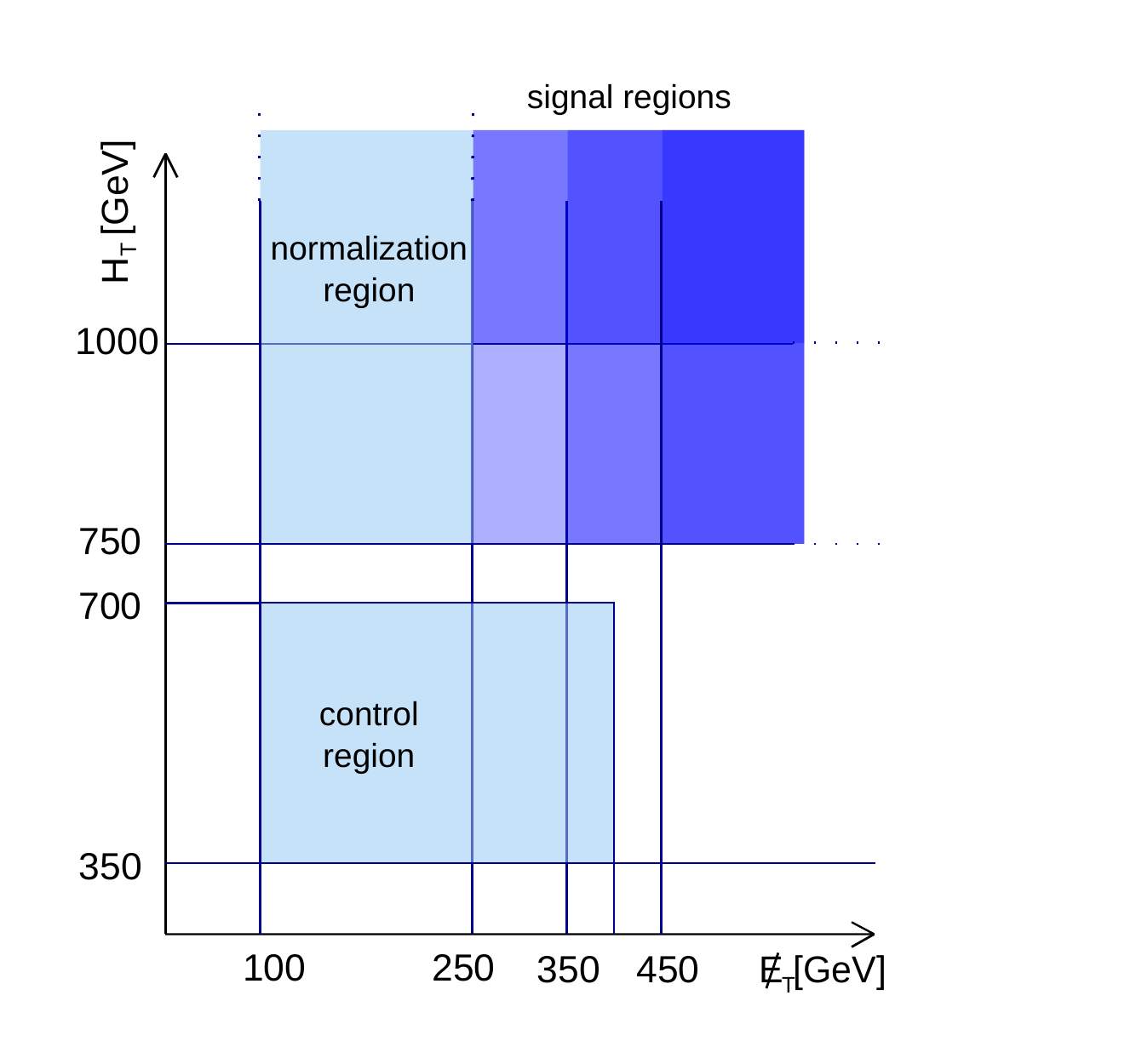}
\caption{
  Graphical representation of the different regions in \HT vs.\ \ETslash space used in the \ETslash~method.
}\label{fig:METTregions}
\end{center}
\end{figure}

\begin{figure*}
\begin{center}
\includegraphics[width=.49\textwidth]{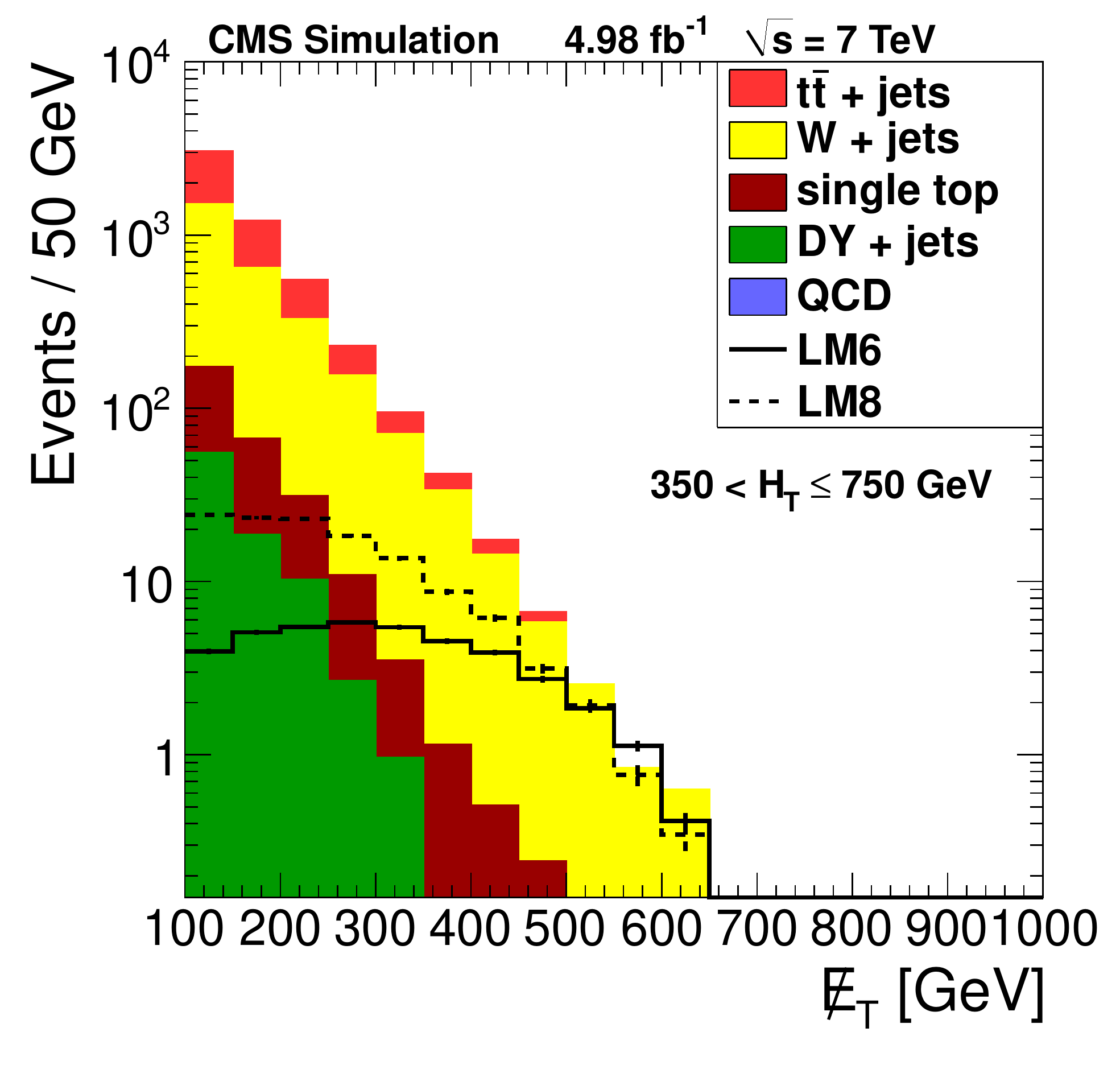}
\includegraphics[width=.49\textwidth]{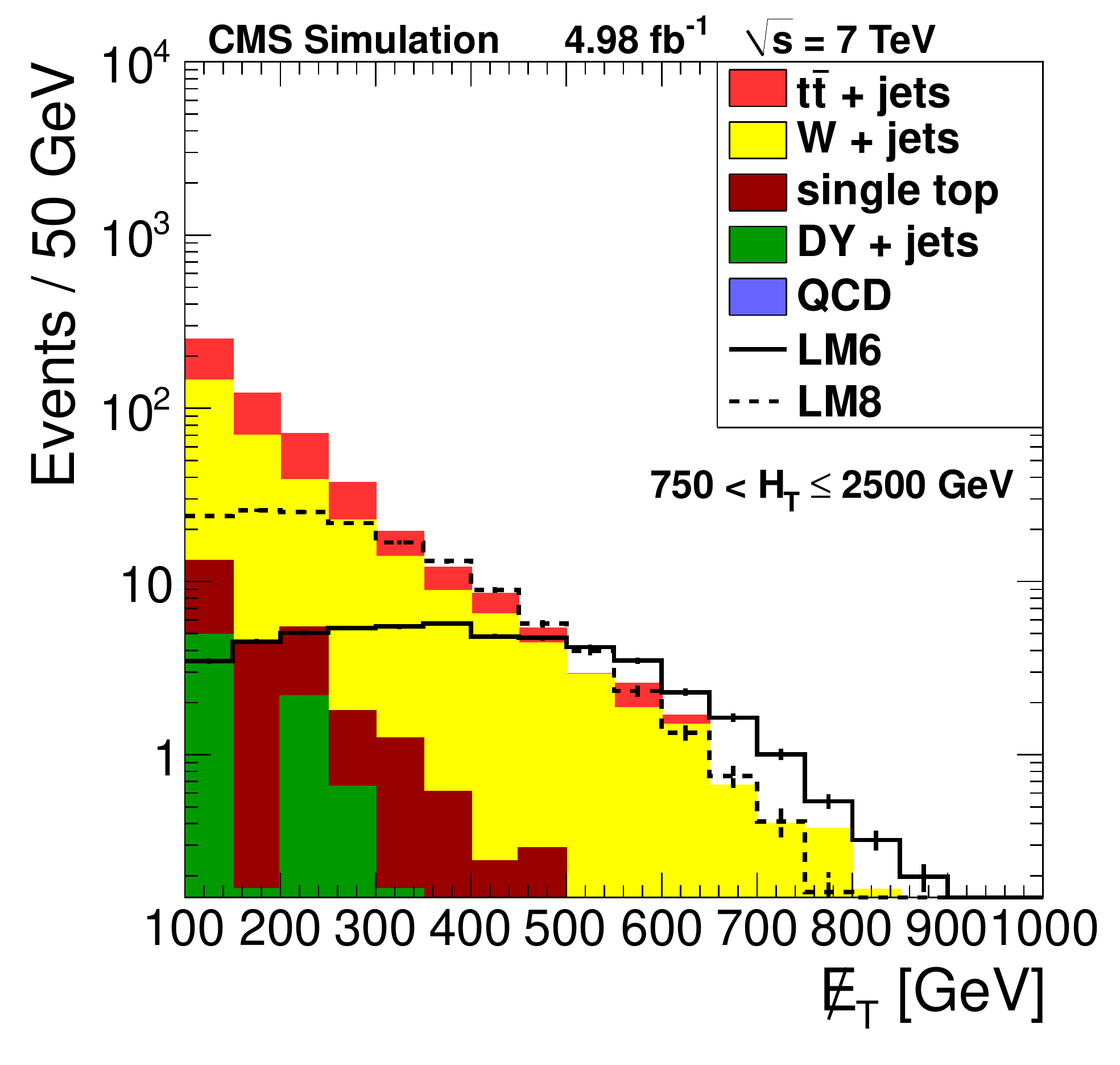}
\caption{
Distribution of \ETslash in the muon channel: simulation of backgrounds and two reference SUSY signals (LM6 and LM8) for (left) $350 < \HT < 750\GeV$ and (right) $750 < \HT < 2500\GeV$. No requirements are imposed on the number of \cPqb\ jets.
}\label{fig:mc-metshape-750}
\end{center}
\end{figure*}

For the \ETslash template method, we consider overlapping signal regions with lower boundaries in \HT at 750\GeV or 1000\GeV, and with lower boundaries in \ETslash at 250\GeV, 350\GeV, and 450\GeV as shown in Fig.~\ref{fig:METTregions}.
All signal regions are restricted to $\HT<2.5\TeV$ and $\ETslash<2\TeV$ since the uncertainties for the prediction increase for very high values of these variables while the additional signal yield is small.
In the \ETslash template approach, parameters of a model for the true \ETslash spectrum are obtained from a fit to a control region in data defined by $350 < \HT < 700\GeV$ and $100 < \ETslash < 400\GeV$.
Separate \ETslash\ models are used for the dominant background processes: \Wmjets, \Wpjets, and \ttbar production.
The absolute scale for the prediction is obtained from a normalization region defined by $750 < \HT < 2500\GeV$ and $100 < \ETslash < 250\GeV$.
Figure~\ref{fig:mc-metshape-750} shows the difference in the \ETslash distributions of the simulated background and the two reference SUSY signals LM6 and LM8 in the muon channel at low and high \HT.
The \ETslash shape used for the predictions in the signal regions is obtained from data and does not depend on the simulated distribution.
Control and normalization regions have been chosen to provide a sufficiently large range in \ETslash for the fit and to limit signal contamination.
The method provides background estimates for events with 0, 1, and ${\geq}2$ identified \cPqb\ jets  in a natural way.

\subsection{Discrimination of \texorpdfstring{\PW}{W} from \texorpdfstring{\ttbar using \cPqb-jet}{t t-bar using b-jet} identification}\label{sec:BTag}

In order to gain sensitivity to the differences between the \ETslash shapes in \Wjets and \ttbar events, we divide the preselected sample into three bins of \cPqb-jet multiplicity, corresponding to different relative proportions of \ttbar and \Wjets events.
Simulation predicts the \zeroTag bin to contain 76\% \Wjets and 19\% \ttbar events, while the \twoTag bin is dominated by \ttbar events (3\% \Wjets versus 90\% \ttbar events).
The \oneTag bin shows intermediate values (20\% \Wjets versus 72\% \ttbar events).
The ratio of \Wpjets to \Wmjets in the sample is predicted to be approximately 3.

The relative fraction of \Wjets and \ttbar events is estimated from data using a template fit for the event fractions in the three \cPqb-jet multiplicity bins.
The templates are extracted from simulation and corrected for the measured differences in \cPqb-quark and light-flavor tagging probabilities between data and simulation.

The evolution of the ratios in the \zeroTag and \oneTag bins as a function of \HT\ is obtained by dividing the \HT distributions of \Wjets and \ttbar events, weighted according to the global \Wjets-to-\ttbar ratio in these bins obtained as described above.
The \HT distribution for \ttbar events is extracted from the \twoTag bin.
The corresponding shape for \Wjets events is obtained by subtracting the \ttbar contribution from the \zeroTag bin according to the measured \ttbar fraction in this bin.
The ratios measured in the data exhibit no significant trend with \HT.

\subsection{The \ETslashSubSection model}\label{sec:MetModel}

In the region well above the \PW\ mass, namely $\ETslash > 100\GeV$, the true \ETslash spectra of the leading backgrounds are characterized by nearly exponential falling shapes.
Small differences can be observed as functions of the production process, \PW\ polarization, and rapidity distributions.
The functional form $x \exp{(-\alpha x^{\beta})}$ with $\beta=0.5$  provides a satisfactory parametrization of the inclusive distributions within each category (\ttbar, \Wpjets, and \Wmjets).
The shapes for \Wpjets and \Wmjets are distinguished from each other using the lepton charge, and separate models are used for the two lepton flavors in order to take into account differences in the acceptance.

The selection in \HT leads to a clear bias in the \ETslash distribution due to the correlation between the transverse momentum of the \PW\ boson and the hadronic activity balancing this momentum.
The shape of the ratio of the \ETslash spectrum after a selection in \HT to the inclusive spectrum can be well described by error functions, $\erf(x; b, c)$, with two free parameters: the \ETslash value where the ratio reaches 50\%, denoted $b$, and the width, denoted $c$.
The evolution of the parameters $b$ and $c$ can be approximated well by linear functions of \HT: $b(\HT) = b_0 + b_1 \HT$ and $c(\HT) = c_0 + c_1 \HT$.
The values for $b_0$, $b_1$, $c_0$, and $c_1$ are obtained from simulation and verified with data.
A second-order polynomial is used as an alternative parametrization in order to assign a systematic uncertainty to the residual non-linearity.

The full \ETslash model for a final-state category (\Wpjets, \Wmjets, or \ttbar) in a single \HT bin $i$ with lower and upper limits $\HT{}_{,i}$ and $\HT{}_{,i+1}$ has the form
\begin{equation}\label{eq:trueMetModel}\begin{split}
  {\cal M}_{i}(x) \sim& x \exp(-\alpha {x}^{0.5})  \times\\
  &(1+\erf(x\,;\,b_0 + b_1 \HT{}_{,i}\,,\,c_0 + c_1 \HT{}_{,i})) \times \\
  & (1-\erf(x\,;\,b_0 + b_1 \HT{}_{,i+1}\,,\,c_0 + c_1 \HT{}_{,i+1})) .
\end{split}\end{equation}

\begin{figure*}
\begin{center}
\includegraphics[width=.47\textwidth]{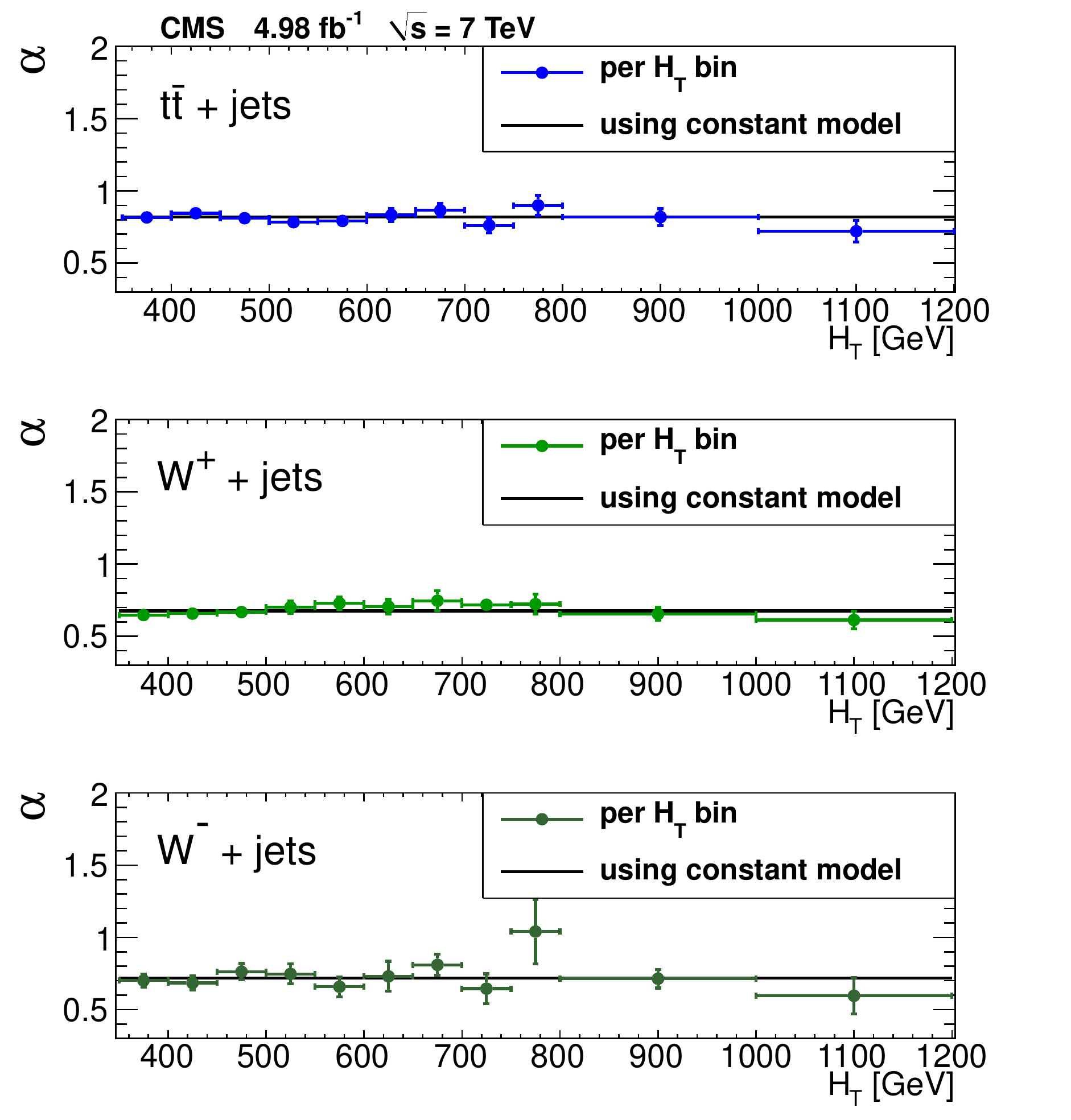}
\includegraphics[width=.47\textwidth]{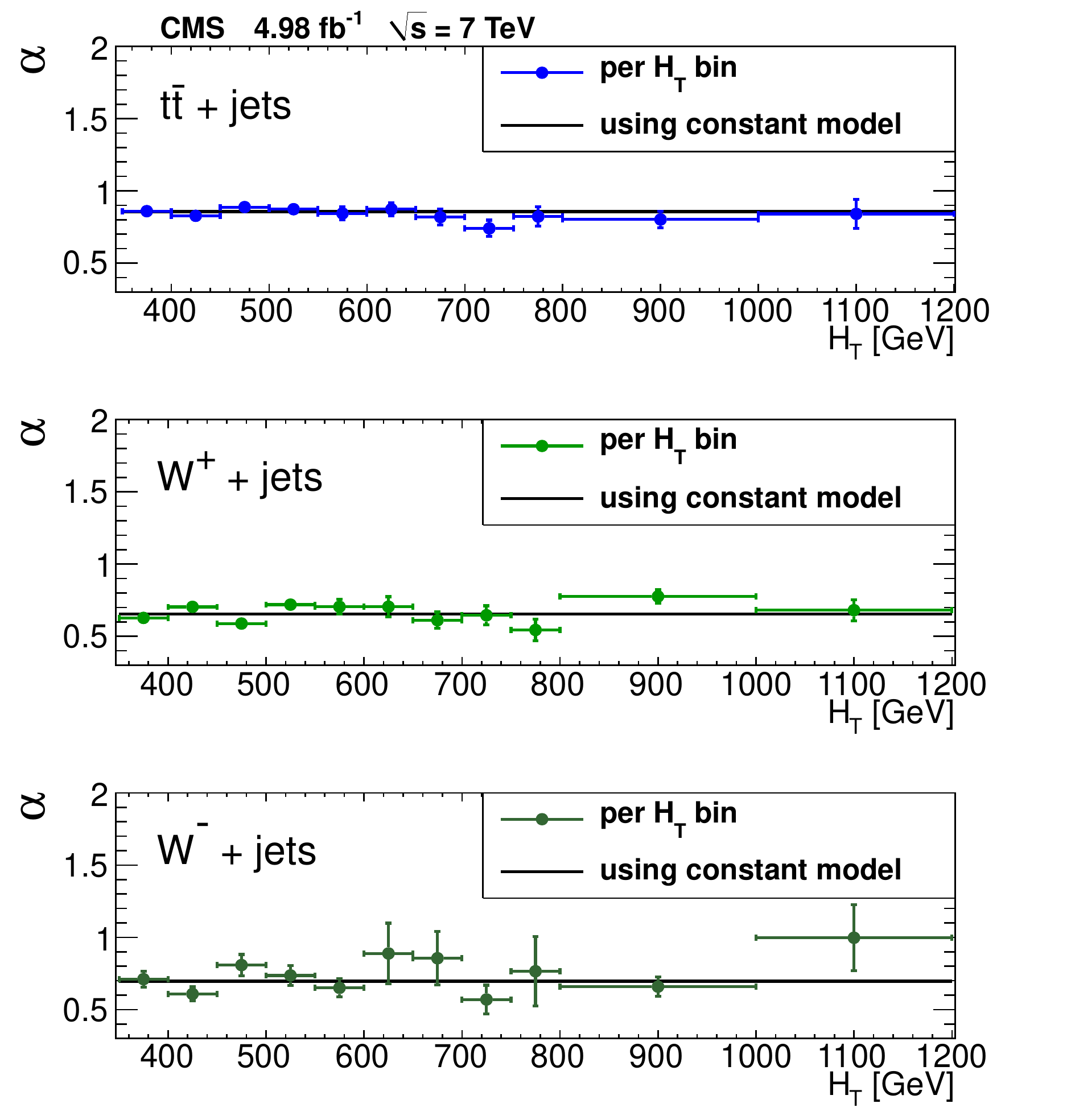}
\caption{
The fitted parameter $\alpha$ as a function of \HT for three subsamples of the (left) muon and (right) electron channel: from top to bottom \ttbar, \PWp+jets, and \PWm+jets are shown.
For the parameters of the error functions a linear dependence on \HT is imposed.
The points show the values of $\alpha$ obtained for individual bins in \HT.
The solid lines correspond to fits to constant values of $\alpha$ in the control regions.
}\label{fig:expVsHT}
\end{center}
\end{figure*}

The categories are combined with the weights described above.
The results of fits to the parameter $\alpha$ in bins of \HT after constraining the parameters $b$ and $c$ to linear functions are shown in Fig.~\ref{fig:expVsHT}.
They show no significant trend, and a single value is used for each category in the final estimate.

As the model for the true \ETslash spectrum is empirical, systematic uncertainties due to the choice of the model have been evaluated by varying the parameter $\beta$ in the exponential form; the parameters $b$ and $c$ of the error function; and the evolution of $\alpha$, $b$, and $c$ with \HT.
Details are given in Section~\ref{sec:Systematics_METT}.

In order to describe the data, the model for the true \ETslash distribution needs to be modified (``smeared'') to account for the finite detector resolution.
The resolution depends on the hadronic activity and on the time-dependent running conditions.
The response function for this \ETslash smearing can be obtained from QCD multijet events, which do not have a significant amount of true \ETslash \cite{Pavlunin}.
A sample dominated by these events is selected using a set of triggers based only on \HT, and the response functions are extracted in bins of \HT, jet multiplicity, and \cPqb-jet multiplicity.
In each \HT bin the shapes for different jet multiplicities are then combined according to the multiplicity distribution observed in the single-lepton dataset.

The convolution of the true \ETslash distribution with the response functions described above assumes that the contribution to \ETslash from missing particles is uncorrelated in direction with the \ETslash contribution arising from jet mismeasurements.
Simulation indicates that the correlation coefficient between these two contributions is only 0.13, and ancillary studies confirm that the uncertainty on the prediction incurred by ignoring the correlation is negligible.

\subsection{Estimation of the \ETslashSubSection spectrum from data}\label{sec:TemplateFit}

The full \ETslash model  described in the previous subsections is used in a simultaneous fit to \HT bins in the subsamples defined by the three \cPqb-jet multiplicities, the two lepton flavors, and the two charges.
The bin sizes in \HT are chosen to ensure adequate data in each bin.
The parameters $\alpha$ resulting from the fits to data and to simulation are summarized in Table~\ref{tab:fittedAlphas}.

The predictions for each of the signal regions are obtained by integrating the \ETslash model in bins of \HT.
In each \HT bin the \ETslash distribution is normalized to the observed number of events at $100 < \ETslash < 250\GeV$.
The final estimate is obtained by summing over all \HT bins.
The statistical uncertainty on the prediction for each signal region is evaluated by pseudo-experiments, repeating the prediction with values for $\alpha$ in the different categories sampled according to the central value and covariance matrix provided by the fit.

In Table~\ref{tab:fitResultsDetails} the predictions of the fit of the \ETslash model to simulated events are compared to the true values for regions defined by lower limits of 750\GeV and 1000\GeV on \HT, and of 250, 350, and 450\GeV on \ETslash.
Good agreement is observed.
The results from data are summarized in Section~\ref{sec:Result}.

\begin{table}[hbt]
\topcaption{Fit results for the parameter $\alpha$ from the control regions in data and simulation.
The quoted uncertainties are statistical.}\label{tab:fittedAlphas}
\scotchrule[l|c||c|c|c]
&  & \Wpjets & \Wmjets & \ttbar \\ \hline
$\mu$ & data & $0.676 \pm 0.014$ & $0.717 \pm 0.024$ & $0.818 \pm 0.014$ \\
      & simulation   & $0.641 \pm 0.019$ & $0.709 \pm 0.024$ & $0.819 \pm 0.013$ \\ \hline
\Pe   & data & $0.655 \pm 0.015$ & $0.697 \pm 0.026$ & $0.857 \pm 0.016$ \\
      & simulation   & $0.651 \pm 0.013$ & $0.736 \pm 0.025$ & $0.806 \pm 0.013$ \\
\donescotchrule
\end{table}

\begin{table}[bth]
\topcaption{Predicted and true event counts in simulation for different signal regions. Uncertainties are statistical.}\label{tab:fitResultsDetails}
\scotchrule[l||rcr|rcr||rcr|rcr]
& \multicolumn{6}{c||}{$750 < \HT < 2500\GeV$} & \multicolumn{6}{c}{$1000 < \HT < 2500\GeV$} \\\hline
& \multicolumn{6}{c|}{simulation} & \multicolumn{6}{c}{simulation} \\
& \multicolumn{3}{c|}{ predicted} & \multicolumn{3}{c||}{ true} & \multicolumn{3}{c|}{ predicted} & \multicolumn{3}{c}{ true}\\
\hline & \multicolumn{12}{c}{$250 < \ETslash < 2000\GeV$ }\\\hline
Total&     196\phantom{.0}  & $\pm$ &  11\phantom{.0} & 183.2  & $\pm$ &    5.1 &  52.0  & $\pm$ &   4.3 &  53.4  & $\pm$ &   2.7\\
$0$ \cPqb\ tag&  129.7  & $\pm$ &   8.6 & 113.4  & $\pm$ &   3.4 &   35.1  & $\pm$ &  3.6 &   31.5  & $\pm$ &   1.8\\
$1$ \cPqb\ tag&   47.4  & $\pm$ &   3.2 &  48.5  & $\pm$ &   3.1 &   11.3  & $\pm$ &   1.5 &  15.9  & $\pm$ &   1.7 \\
${\ge}2$ \cPqb\ tags&   19.3  & $\pm$ &   1.9 &  21.2  & $\pm$ &   2.2 &    5.7  & $\pm$ &   1.0 &   6.0  & $\pm$ &   1.9 \\
\hline & \multicolumn{12}{c}{$350 < \ETslash < 2000\GeV$ }\\\hline
Total&     74.5  & $\pm$ &    5.1 &  71.9  & $\pm$ &   2.9 &21.9  & $\pm$ &   2.2 &  23.3  & $\pm$ &   1.7\\
$0$ \cPqb\ tag&   52.8  & $\pm$ &   4.4 &  48.1  & $\pm$ &   2.0 &15.7  & $\pm$ &  1.8 &  13.6  & $\pm$ &    1.0\\
$1$ \cPqb\ tag&   16.2  & $\pm$ &   1.2 &  17.2  & $\pm$ &   1.7 & 4.3  & $\pm$ &    0.6 &   6.7  & $\pm$ &   1.1\\
${\ge}2$ \cPqb\ tags&    5.6  & $\pm$ &   0.6 &   6.7  & $\pm$ &   1.3 & 1.9  & $\pm$ &   0.3 &   3.0  & $\pm$ &   0.9 \\
\hline & \multicolumn{12}{c}{$450 < \ETslash < 2000\GeV$ }\\\hline
Total&     28.1  & $\pm$ &   2.4 &  30.2  & $\pm$ &   1.8 &   9.5  & $\pm$ &   1.1 &  11.2  & $\pm$ &    1.1\\
$0$ \cPqb\ tag&   21.0  & $\pm$ &   2.1 &  21.1  & $\pm$ &   1.2 &  7.2  & $\pm$ &  1.0 &   7.2  & $\pm$ &   0.7 \\
$1$ \cPqb\ tag&    5.5  & $\pm$ &   0.5 &   6.4  & $\pm$ &   0.9 &  1.7  & $\pm$ &   0.3 &   2.6  & $\pm$ &   0.6 \\
${\ge}2$ \cPqb\ tags&    1.6  & $\pm$ &   0.2 &   2.7  & $\pm$ &   0.8 &  0.6  & $\pm$ &   0.1 &    1.4  & $\pm$ &   0.6 \\
\donescotchrule
\end{table}

\subsection{Experimental systematic uncertainties}
\label{sec:Systematics_experimentalMETT}

The results can be affected by systematic uncertainties, which arise from detector effects, assumptions made about the shape of the distribution, theoretical uncertainties, and the contamination due to other backgrounds.
The impact of these uncertainties on the prediction can be quantified by a relative variation defined as
$\delta\rho = (N'_\text{pred}/N'_\text{true}) / (N_\text{pred}/N_\text{true})-1$
where  $N_\text{pred}$ ($N_\text{true}$) is the predicted (true) number of events and the prime denotes the values with the systematic effect included.
For those uncertainties that only affect the estimation procedure but not the true number of events in the signal region, this amounts to the relative change in the prediction.
For all other sources, $\delta\rho$ determines the variation in closure estimated with simulation, i.e., how well the prediction follows the change of events in the signal region.

Miscalibration of the jet energy scale (JES) leads to a modification of the true number of events in the signal region but is compensated to a large extent by a corresponding change in the predicted number of events.
The effect due to the uncertainty on the JES is determined by shifting the energy of jets with $\pt>10\GeV$ and $|\eta|<4.7$
in simulated events up and down according to $\pt$- and $\eta$-dependent uncertainties that have been measured using dijet and $\gamma$/\zpj
events~\cite{Chatrchyan:2011ds}.
The applied shifts, which are 1--3\% for jets with $\pt>40\GeV$ and $|\eta|<2.0$ and increase towards lower
$\pt$ and higher $|\eta|$, are propagated to the \ETslash result.
The uncertainty on the energy of jets with $\pt<10\GeV$, referred to as unclustered
energy, is assumed to be 10\%. This uncertainty is also propagated to the \ETslash result assuming full correlation with the JES uncertainty.
For the muon channel and a signal region inclusive in \cPqb-jet multiplicity and defined by $\HT>1000\GeV$ and $\ETslash > 250\GeV$, the variations are $+14\%$ and $-30\%$, respectively, while the systematic uncertainty $\delta\rho$ is $6\%$.

Lepton efficiencies are expected to have a small impact on the background prediction, because an overall change of scale is compensated by a corresponding change in the normalization regions, and the preselection cuts have been chosen to use only kinematic regions with stable trigger and reconstruction efficiencies.
Therefore only small changes are expected in the ratios of yields between the signal and the normalization regions.
In order to test the impact of a possible non-uniformity, the lepton efficiencies are lowered by 5\% in the endcap regions and by a linear variation of $-20\%$ to 0\% in the low \pt range of 20 to 40\GeV, where any residual effect of the efficiency in the threshold region would have the highest impact.

Over the course of the data collection period, the maximum instantaneous luminosity per bunch crossing
and, hence, the average number of simultaneous collisions, changed dramatically. Simulated
events are matched to the pileup conditions observed in data using the distribution
of the number of reconstructed primary vertices, and the simulation provides a satisfactory description of the dependence of several key observables as a function of the number of simultaneous collisions.
Possible residual effects are tested by varying the event weight according to the reconstructed number of primary vertices $n_\text{vtx}$ by ${\pm}5\% \times (n_\text{vtx} - \langle n_\text{vtx}\rangle)$ around the mean number $\langle n_\text{vtx}\rangle = 7$.

Differences between the efficiencies to tag \cPqb-quark, \cPqc-quark, and light-flavor jets in data and simulation are taken into account by applying $\pt$- and $\eta$-dependent scale factors to the simulated events.
These scale factors are measured in data using QCD multijet event samples with uncertainties on the order of a few percent \cite{ref:PAS-BTV-11-004}.
Variations in the efficiency and purity of the \cPqb-jet identification would move events among the three \cPqb-tag multiplicity bins and change the fractions of \Wjets and \ttbar events in each bin.
The size of this effect is estimated by varying efficiencies and mistagging rates within the uncertainties.
As expected, the determination of the fractions based on fits to the \cPqb-jet multiplicity compensates for these changes and the residual effects are small.

\subsection{Model-related systematic uncertainties for the \ETslashSubSection templates method}
\label{sec:Systematics_METT}

The background estimation procedure is designed to provide individual estimates of the \ETslash distribution of each of the leading backgrounds: \ttbar, \Wpjets, and \Wmjets.
The accuracy of the separation between \ttbar and \Wjets events is tested by varying the \ttbar and \Wjets cross sections individually by one third.
Moreover, the sensitivity of the fit results to the \cPqb-jet multiplicity distribution is estimated by varying the resulting ratio of \Wjets to \ttbar events by its uncertainty.
The corresponding effect is small.
The impact of other background sources, in particular of the contribution from dilepton events, is tested by varying the amount of all non-leading backgrounds by ${\pm}50\%$.

The uncertainty on the \ETslash model is tested by varying the $\beta$ parameter by $\pm 10\%$ with respect to its nominal value of 0.5. This variation is motivated by the uncertainty from fitting $\beta$ in single-lepton events with two jets.
As shown in Fig.~\ref{fig:expVsHT}, the parameter $\alpha$ shows no significant dependence on \HT.
The uncertainty on this assumption is quantified by imposing a slope according to the uncertainties of the linear fits as a function of \HT in the control region.
These two model-related effects constitute the dominant systematic uncertainties in the background estimation.
For the parameters of the error functions $b$ and $c$, 16 independent variations are considered in the eigenbasis of the parameters of the linear functions describing the evolution in \HT, and the model describing this evolution is changed from linear to quadratic.
The effect of these variations is rather small, since the prediction for any signal region is a sum of many \HT bins, and the variations of the error function parameters tend to cancel each other.

An additional source of uncertainty is due to the \PW\ polarization, which would alter the \ETslash distribution for a given momentum of the \PW\ boson.
In order to quantify this uncertainty we modify the generator-level polarization distributions in bins of lepton \pt and rapidity according to varied scenarios.
The fit is performed for each of the modified datasets, and the highest $\delta\rho$ is then assigned as a systematic uncertainty.

The systematic uncertainties for the signal region defined by $\HT > 1000\GeV$ and $\ETslash > 250\GeV$  are presented in Tables~\ref{tab:BkgSyst_1000-250} and \ref{tab:BkgSyst_BTag}.
Table~\ref{tab:BkgSyst_1000-250} contains all contributions that are not directly related to \cPqb-jet identification.
They have been evaluated in a \Wjets and a \ttbar dominated subsample, defined as events without or with at least one identified \cPqb\ jet, respectively.
Table~\ref{tab:BkgSyst_BTag} lists the \cPqb-tagging related systematic effects in the three \cPqb-jet multiplicity bins.

\begin{table*}[hbt]
\topcaption{
Relative systematic uncertainties ($\delta\rho$) not directly related to \cPqb\ tagging for the background estimation in the signal region $1000 < \HT < 2500\GeV$ and $250 < \ETslash < 2000\GeV$.
}\label{tab:BkgSyst_1000-250}
\scotchrule[l||r|r|r||r|r|r]
& \multicolumn{3}{c||}{$\mu$ channel} & \multicolumn{3}{c}{$\Pe$ channel}\\ \hline
Source & Total & $0$ \cPqb\ tag & ${\ge}1$ \cPqb\ tag& Total & $0$ \cPqb\ tag  &${\ge}1$ \cPqb\ tag \\ \hline
Jet and \ETslash scale &     6.0\% &    7.5\% &    7.2\% &    3.1\% &    5.6\% &    2.1\% \\
Lepton efficiency &     0.4\% &    0.3\% &    0.6\% &    0.6\% &    1.3\% &    0.7\% \\
Pileup &     0.1\% &    0.1\% &    0.2\% &    0.3\% &    1.5\% &    0.4\% \\
$\PW$\ polarization &     0.5\% &    0.6\% &    0.1\% &    1.3\% &    1.8\% &    0.3\% \\
Non-leading backgrounds &     0.7\% &    0.4\% &    0.4\% &    4.0\% &    3.0\% &    6.2\% \\
Dilepton contribution &     0.1\% &    0.5\% &    0.7\% &    0.6\% &    1.2\% &    0.6\% \\
$\sigma(\,$\ttbar$)$ &     1.2\% &    2.3\% &    1.6\% &    0.7\% &    1.8\% &    2.0\% \\
$\sigma(\,$\Wjets$)$ &     1.3\% &    2.9\% &    2.3\% &    2.6\% &    1.6\% &    2.8\% \\
Exponent $\beta$\ \ttbar  &     1.6\% &    0.2\% &    5.3\% &    1.8\% &    0.3\% &    4.8\% \\
Exponent $\beta$\ \Wpjets &     3.5\% &    4.4\% &    1.3\% &    3.6\% &    4.6\% &    1.5\% \\
Exponent $\beta$\ \Wmjets &     0.7\% &    0.8\% &    0.3\% &    0.9\% &    1.4\% &    0.9\% \\
$\alpha$ slope \ttbar &    11.0\% &    2.4\% &   29.3\% &   14.8\% &    5.0\% &   34.3\% \\
$\alpha$ slope \Wpjets &    15.9\% &   20.6\% &    6.0\% &   16.5\% &   22.2\% &    5.1\% \\
$\alpha$ slope \Wmjets &     4.9\% &    8.2\% &    2.0\% &    5.6\% &    8.7\% &    0.5\% \\
Error function parameters &     4.1\% &    4.6\% &    2.9\% &    3.1\% &    3.2\% &    2.7\% \\
\donescotchrule
\end{table*}

\begin{table*}[hbt]
\topcaption{
Relative systematic uncertainties related to \cPqb\ tagging in the signal region $1000 < \HT < 2500\GeV$ and $250 < \ETslash < 2000\GeV$.
}\label{tab:BkgSyst_BTag}
\scotchrule[l||r|r|r|r|r]
\hline
Source & Total & $0$ \cPqb\ tag & $1$ \cPqb\ tag & ${\ge}1$ \cPqb\ tag& ${\ge}2$ \cPqb\ tags \\ \hline
& \multicolumn{5}{c}{$\mu$ channel} \\ \hline
\Wjets/\ttbar ratio  &     2.9\% &    2.1\% &    6.1\% &    4.8\% &    2.4\% \\
\cPqb-tagging efficiency &     2.0\% &    1.5\% &    2.2\% &    1.3\% &    5.1\% \\
Mistagging rate &           0.4\% &    0.4\% &    0.7\% &    0.9\% &    0.6\% \\ \hline
& \multicolumn{5}{c}{$\Pe$ channel} \\ \hline
\Wjets/\ttbar ratio  & 1.1\% &    2.4\% &    2.6\% &    2.3\% &    2.3\% \\
\cPqb-tagging efficiency & 2.2\% &    1.6\% &    0.8\% &    1.7\% &    3.6\% \\
Mistagging rate           & 0.3\% &  0.4\% &    0.4\% &    0.2\% &    0.1\% \\
\donescotchrule
\end{table*}

In simulated event samples, the background estimation procedure produces results that are compatible with the simulated rates.
Conservatively, a systematic uncertainty using the maximum of the statistical uncertainty of this comparison and of the absolute value of the deviation is assigned.
For the signal region mentioned above, this amounts to 5.9\%.
We also evaluate the effect of possible differences in the \ETslash distributions between the different \cPqb-tag bins.
In order to test the sensitivity to possible deviations in the low-\HT control region used for the fit, we have evaluated the relative variations in the predictions for the \oneTag (\twoTag) bin in data by repeating the fit without the \twoTag (\oneTag) bin.
The uncertainty in the \geqOneTag bin was set to the average of the variations in the \oneTag and \twoTag bins.
For the signal region mentioned above, the uncertainties are 2.0\%, 4.2\%, and 8.5\% for the \oneTag, \geqOneTag, and \twoTag bins, respectively.

\section{The Factorization Method}
\label{sec:SR_BG_ABCD}

\begin{figure}[bhtp]
 \begin{center}
  \includegraphics[width=0.45\textwidth]{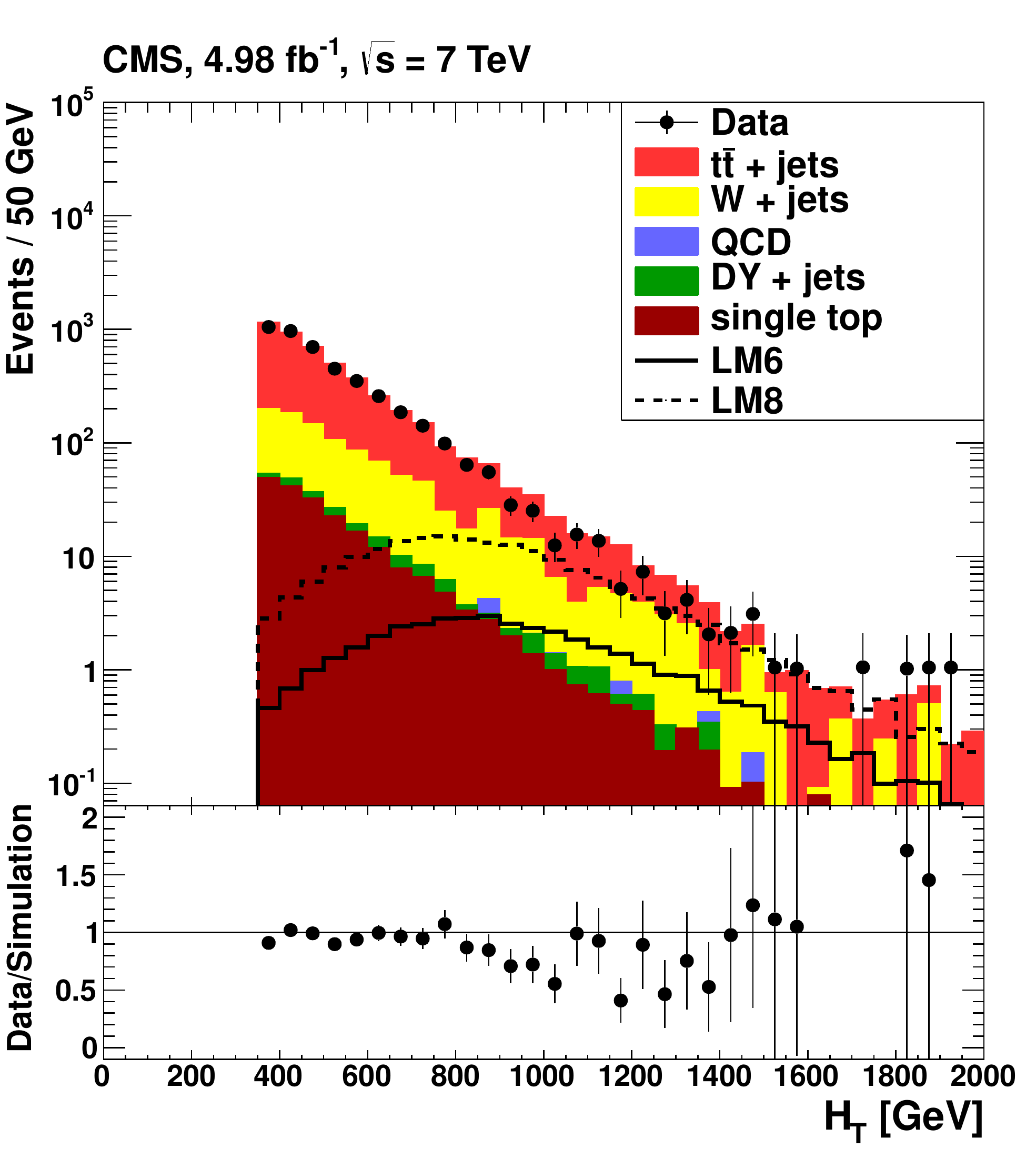}
 \includegraphics[width=0.45\textwidth]{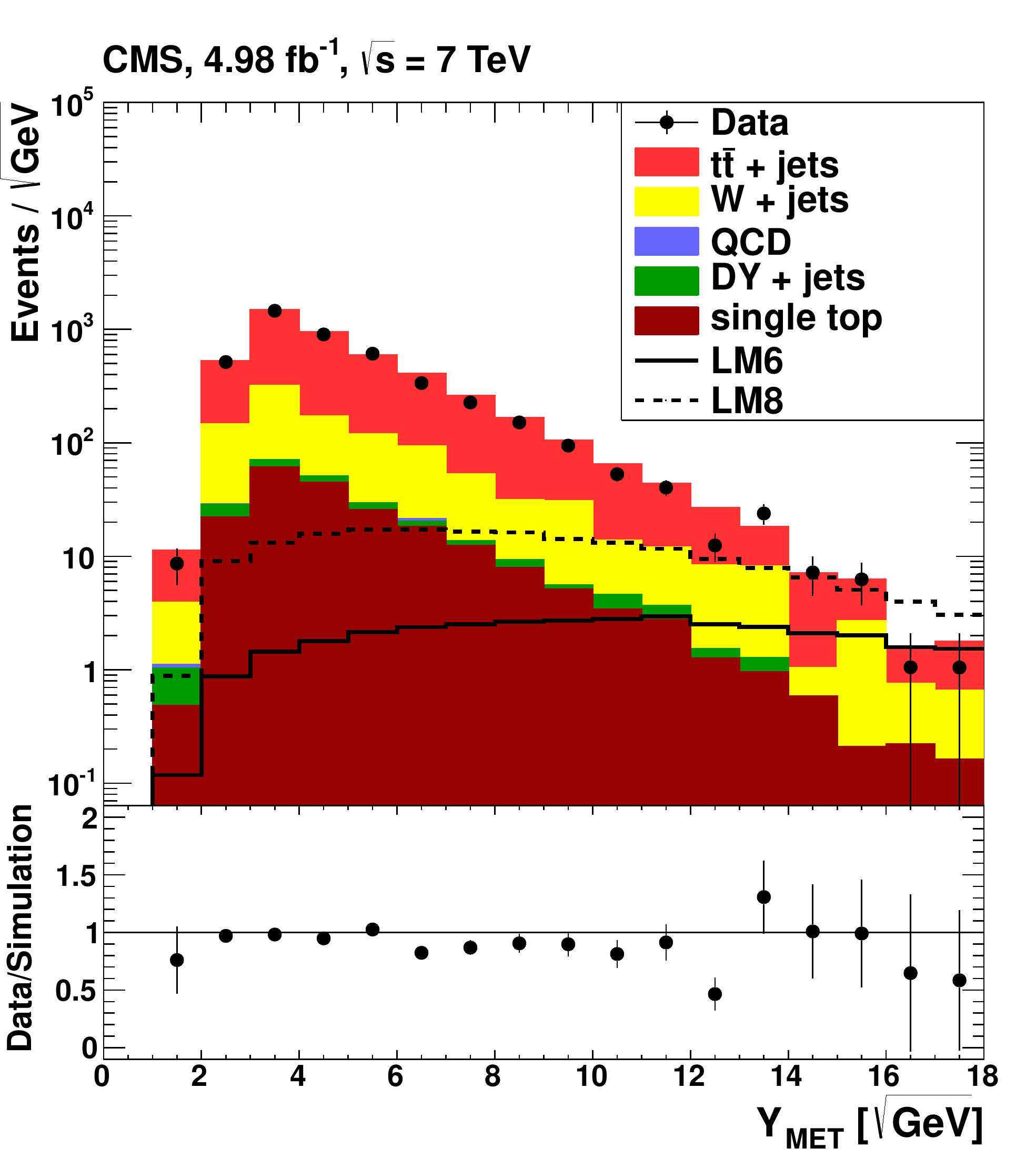}
 \end{center}
 \caption{Distributions of (\cmsLeft) $\HT$ and (\cmsRight) $Y_\mathrm{MET}$ for data compared to the
   different SM processes. The muon and electron channels are combined and at least one b tag is required.
The CMS data are represented by solid points and the simulated SM events by stacked histograms. The
two lines represent possible signal scenarios. The simulation is normalized
to the integrated luminosity of the data sample.}\label{fig:HT_Y}
\end{figure}

The factorization method is based on the variables $\HT$ and $Y_\mathrm{MET}$, which are shown for the inclusive
1 b-tag selection for data and simulated SM events in Fig.~\ref{fig:HT_Y}. The SM simulation lies systematically above the data,
showing the need for background estimation from data. Since $\HT$ and $Y_\mathrm{MET}$ are nearly uncorrelated for \ttbar production,
which constitutes the main background in events with at least one \cPqb\ jet,
a factorization ansatz in the $Y_\mathrm{MET}$--$\HT$ plane can be used to estimate the background
contribution, namely from control regions with low $\HT$ and/or $Y_\mathrm{MET}$.

For the factorization method, a minimum of $\HT>375$\GeV and $\ETslash>60$\GeV is required together with at least four jets with
$\pt>40$ GeV. For a precise
estimation of the number of background events in the signal region, it is essential to have enough events in the
control regions. Therefore, the definition of the signal region depends on the number of required b tags. The
analysis is performed, and results are presented, in three channels according to the number of b tags: 1, 2,
and ${\ge}3$ b tags, selected with the track-counting algorithm. In addition we
study the 0 b-tag bin for cross checks and use a combined ${\ge}1$
b-tags bin for limit setting in the CMSSM case.
The signal region is defined as $\HT>800$\GeV and $Y_\mathrm{MET} > 5.5\,\sqrt\GeVns{}$ for the 1, the 2,
and the combined ${\ge}1$ b-tag bins, and $\HT>600$\GeV and $Y_\mathrm{MET} >
6.5\,\sqrt\GeVns{}$ for the 0 and ${\ge}3$ b-tag bins. These regions are optimized to balance two opposing requirements: a small background contribution to the signal region but nonetheless enough
background events in the three control regions that the statistical uncertainties on the background
predictions are small.

The signal region is populated with events described by the tails of SM distributions and mismeasurement related to the finite detector resolution.
\begin{figure}[bht]
 \begin{center}
 \includegraphics[width=0.49\textwidth]{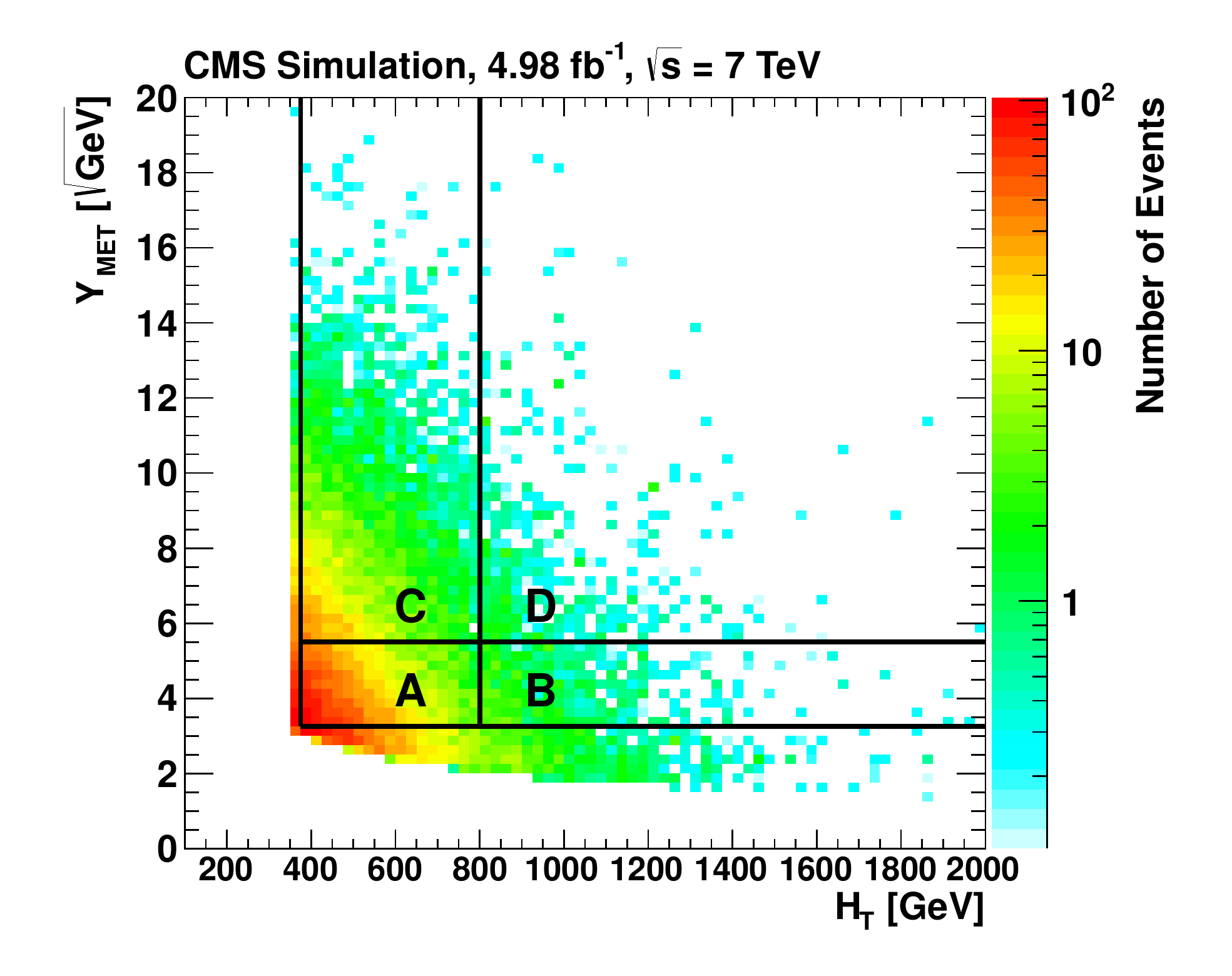}
 \includegraphics[width=0.49\textwidth]{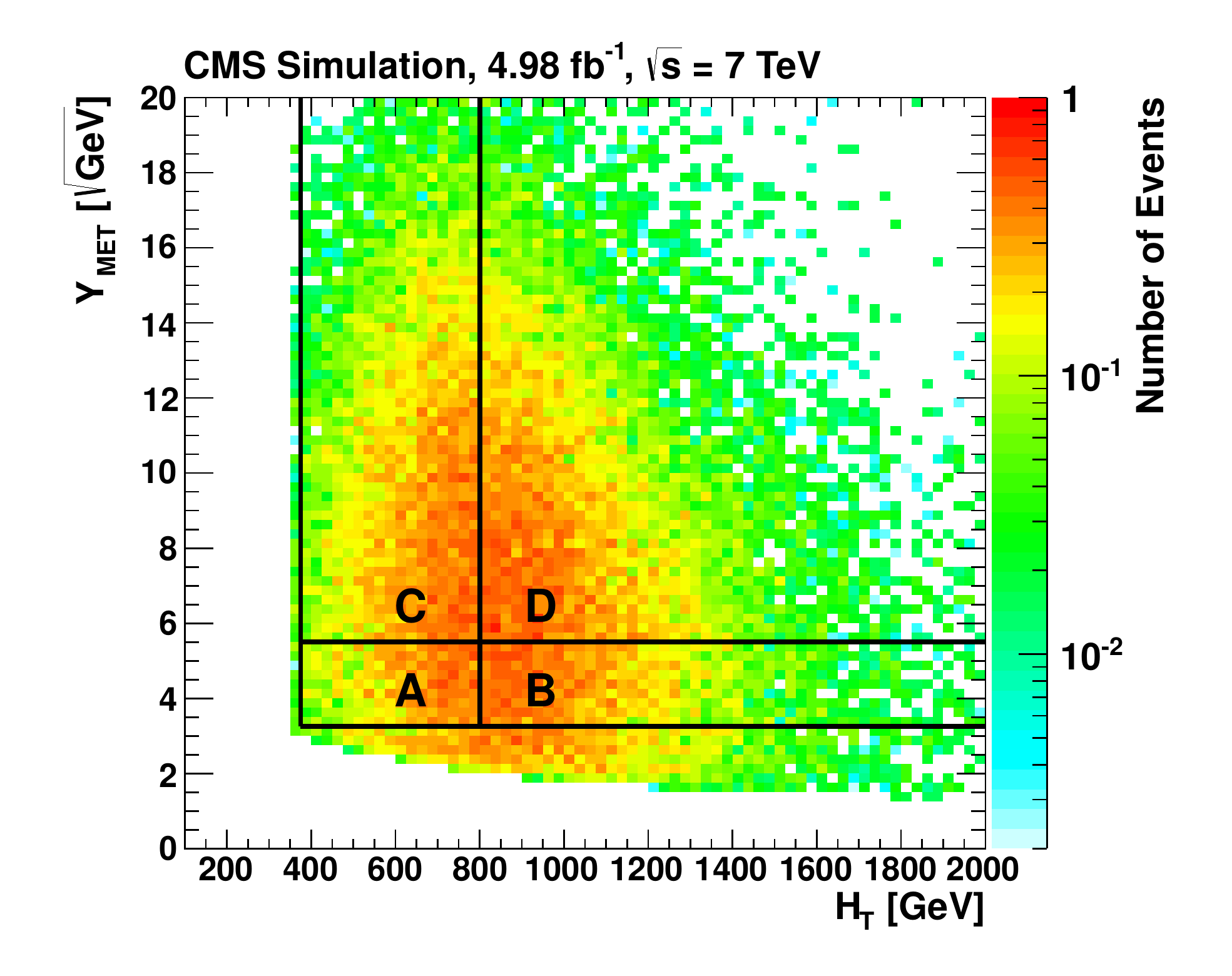}
 \end{center}
 \caption{Distributions of $Y_\mathrm{MET}$ vs.\ $\HT$ for (\cmsLeft) the SM background and (\cmsRight) the SUSY LM8 scenario. The muon and electron channels are combined and at least one b tag is required.}\label{fig:ABCD}
\end{figure}
The control regions ($A$, $B$, and $C$) and the signal region ($D$) used for the factorization method with $\HT$ and $Y_\mathrm{MET}$
are defined in Table~\ref{tab:regions}.
\begin{table}[tbh]
\topcaption{Definition of the different regions used for the
  factorization method with $\HT$ and
  $Y_\mathrm{MET}$. Two sets of selections are defined depending on the number of b tags.
  Region $D$ is expected to be signal dominated. }\label{tab:regions}
\scotchrule[c|c|c||c|c]
\multirow{2}{*}{Region}
   & \multicolumn{2}{c||}{\rule{0pt}{1.05em}b tags: 1,\,2,\,${\ge}1$} &  \multicolumn{2}{c}{b tags: 0,\,${\ge}3$} \\
\cline{2-5}
   &  \rule{0pt}{1.2em} $\HT$/GeV & $Y_\mathrm{MET}$/$\sqrt\GeVns{}$ &  $\HT$/GeV & $Y_\mathrm{MET}$/$\sqrt\GeVns{}$ \\
\hline
$A$  & \rule{0pt}{1.1em} $375 - 800$ & \multirow{2}{*}{$3.25 - 5.5$} & $375 - 600$ & \multirow{2}{*}{$3.25 - 6.5$} \\
$B$  &                   $>800$      &                                           & $>600$      &                               \\
\cline{2-5}
$C$  & \rule{0pt}{1.1em} $375 - 800$ & \multirow{2}{*}{$>5.5$}       & $375 - 600$ & \multirow{2}{*}{$>6.5$}  \\
$D$  &                   $>800$      &                                           & $>600$      &                          \\
\donescotchrule
\end{table}

The number of background events $\hat{N}_D$ in region $D$ is estimated from the three control regions as:

\begin{equation} \label{eqn:kappa}
\hat{N}_D= \kappa N_B \;\frac{N_C}{N_A}\,.
\end{equation}

Were the two variables completely uncorrelated, the correlation factor $\kappa$ would equal one. As $Y_\mathrm{MET}$ and $\HT$ have a small
correlation, the factor $\kappa$ is determined to be 1.20 with an overall uncertainty of about 11\%, as discussed in Section~\ref{sec:Systematics_ABCD}.

The distribution of SM events in the $Y_\mathrm{MET}$--$\HT$ plane after the event selection in the combined muon and electron
channel with the requirement of at least one b tag is presented in Fig.~\ref{fig:ABCD}(a). The corresponding results for the LM8 SUSY scenario
are presented in Fig.~\ref{fig:ABCD}(b). It is observed that the SM events are
mainly located in the control regions, while the signal events are present in the signal and the control regions. The signal contamination
is taken into account in the likelihood model for the scans during limit setting.

\subsection{Systematic uncertainties for the factorization method}
\label{sec:Systematics_ABCD}

As for the \ETslash template method, many systematic effects result
 in small uncertainties only, since the background prediction is affected
 in the same way as the measurements.

Values of $\kappa$ as defined in Eq.~(\ref{eqn:kappa}) for the main SM background processes are
shown in Table~\ref{tab:kappa_for_SM_1} for both signal region definitions and different numbers of b tags.
For the dominant background, due to \ttbar events,  as well as for the backgrounds from single-top and W+jets events,
the correlations are larger than one, indicating a residual correlation.
Besides these processes we expect only small contributions from \zpj events.
The stability of the correlation factor $\kappa$ has been
tested extensively, and the observed correlation is accounted for by the value of $\kappa$ from simulation.
To account for uncertainties in the cross sections of the main SM processes, each cross section is scaled
up and down by 50\%, and the corresponding uncertainty on $\kappa$ is determined.

\begin{table*}[htb]
\topcaption{
Correlation factor $\kappa$ between $\HT$ and $Y_\mathrm{MET}$ for the main SM background processes and a for different
number of b tags, for the two signal regions. For purposes of illustration, the corresponding
results for a sample with 0 b tags is also shown. While the 0 b-tag
sample is dominated by \wpj events, the channels that include b tags contain mainly \ttbar
events. Only statistical uncertainties are shown.
} \label{tab:kappa_for_SM_1}
\small
\renewcommand{\arraystretch}{1.2}
\scotchrule[l|l|l|l|l|l]
Signal region & No. of b tags &$\kappa$ (\ttbar)& $\kappa$ (single top)& $\kappa$ (W+jets) & $\kappa$ (all SM) \\
\hline
\multirow{3}{*}{\parbox{10em}{\centering $\HT > 800$\GeV \\ \rule{0pt}{1.1em} $Y_\mathrm{MET}>5.5$~$\sqrt{\GeVns{}}$ }}
& \phantom{$\ge\,$}1 b-tag\phantom{s} & 1.16 $\pm$ 0.02 & 1.14 $\pm$ 0.14 & 1.17 $\pm$ 0.05 & 1.19 $\pm$ 0.03 \\
& \phantom{$\ge\,$}2 b-tags           & 1.22 $\pm$ 0.02 & 1.25 $\pm$ 0.16 & 1.24 $\pm$ 0.10 & 1.23 $\pm$ 0.02 \\
&             $\ge 1$ b-tags          & 1.18 $\pm$ 0.01 & 1.18 $\pm$ 0.10 & 1.18 $\pm$ 0.04 & 1.20 $\pm$ 0.02 \\
\hline \hline
\multirow{2}{*}{\parbox{10em}{\centering $H_T > 600$\GeV \\ \rule{0pt}{1.1em} $Y_\mathrm{MET}>6.5$~$\sqrt{\GeVns{}}$ }}
& \phantom{$\ge\,$}0 b-tags           & 1.14 $\pm$ 0.06 & 1.44 $\pm$ 0.49 & 1.25 $\pm$ 0.04 & 1.25 $\pm$ 0.04 \\
&             $\ge 3$ b-tags          & 1.17 $\pm$ 0.02 & 1.40 $\pm$ 0.18 & 1.24 $\pm$ 0.19 & 1.19 $\pm$ 0.02 \\
\donescotchrule
\end{table*}

Except for the $\ETslash$ requirement, the offline selection criteria are designed to be well above the trigger
thresholds, where the efficiency reaches a plateau. For events with $\ETslash<80\GeV$, the efficiency of the triggers with a
$\HTslash^{\text{trigger}}$ threshold of 40\GeV can be as low as around 80\%. In these cases the prediction is corrected
to account for the inefficiencies.

As the studies above are based on simulation, a cross-check is performed with data in
the 0 b-tag channel, which can be considered as signal-free, since previous analyses have already excluded
this part of phase space \cite{Chatrchyan:2012mfa}.
From this channel a value of
$\kappa = 1.19 \pm 0.13$
is observed in data, while for the SM simulation a value of
$1.25 \pm 0.04$
is extracted. Although the values are consistent within their statistical uncertainties,
a smaller value of $\kappa$ cannot be excluded. We account for this possibility by including an additional
systematic uncertainty of 10\% on the value of $\kappa$. The
uncertainties for the different selections are described in Section~\ref{sec:Systematics_experimentalMETT} and summarized in
Table~\ref{tab:kappa_sys}. The statistical uncertainty in simulation is relatively small, as the b tagging is applied in the simulation by event weights.
In addition, the simulated jet energy resolution (JER)~\cite{Chatrchyan:2011ds} of jets
with $\pt>10$\GeV and $|\eta|<4.7$ is globally increased by 10\% to provide a more realistic description of the data.
The uncertainty on the jet energy resolution is then determined by variation of the corrected simulated JER up and down by ${\pm}10\%$, and propagated
to $\ETslash$.

\begin{table*}[htb]
\topcaption{Overview of the uncertainties on the correlation factor
  $\kappa$ for the different b-tag selections. The signal regions corresponding to the number of required b tags are as defined in Table~\ref{tab:regions}.
All systematic uncertainties are added in quadrature.
The variations in JES, JER, $\pt^\text{lepton}$, and unclustered energy are
propagated to the $\ETslash$. The row labeled '0 b tag' addresses the difference between the values of $\kappa$ in data and simulation.
 \label{tab:kappa_sys}}
\scotchrule[l||c|c|c|c||c]
Variation & $\Delta \kappa$ & $\Delta \kappa$ & $\Delta \kappa$ & $\Delta \kappa$ & $\Delta \kappa$ \\
& (0 b tags) & (1 b tag) & (2 b tags) & (${\ge}3$ b tags)& (${\ge}1$ b tag)\\
 \hline %\hline
JES  & 2.0\%  & 2.7\%  & 1.3\%  & 0.4\%  & 2.0\% \\ %\hline
JER  & 1.1\%  & 2.1\%  & 3.0\%  & 1.5\%  & 2.4\% \\ %\hline
$\pt^\text{lepton}$   & 1.2\%  & 1.5\%  & 1.7\%  & 1.2\%  & 1.6\% \\ %\hline
Unclustered energy  & 0.5\%  & 0.5\%  & 1.1\%  & 0.4\%  & 0.8\% \\ %\hline
Pileup  & 0.7\%  & 0.6\%  & 0.8\%  & 1.9\%  & 0.7\% \\ %\hline
b-tagging scale factor  & 0.1\%  & 0.2\%  & 0.3\%  & 0.3\%  & $<0.1$\% \\ %\hline
Mistagging scale factor  & 0.1\%  & 0.1\%  & 0.2\%  & 0.2\%  & $<0.1$\% \\ %\hline
Cross section variation  & 3.4\%  & 1.0\%  & 2.0\%  & 1.4\%  & 0.4\% \\ %\hline
0 b tag  & 10.0\%  & 10.0\%  & 10.0\%  & 10.0\%  & 10.0\% \\ \hline
\hline
Total uncertainty   & 10.9\%  & 10.7\%  & 10.9\%  & 10.2\%  & 10.7\% \\ \hline %\hline
Statistical uncertainty   & 3.8\%  & 3.7\%  & 2.5\%  & 2.1\%  & 2.3\% \\
\donescotchrule
\end{table*}

Since the value of $\kappa$ is found to be consistent for all channels within the statistical uncertainties, we use the value
$\kappa = 1.20 \pm 0.02$ (stat) found for simulated events with ${\ge}1$ b-tag to describe all channels.
The corresponding systematic uncertainty for each channel is taken from Table~\ref{tab:kappa_sys}. The sum of the statistical and
systematic uncertainty on $\kappa$ corresponds to the systematic uncertainty for the prediction $\hat{N}_D$.

For the comparison of data with simulation, the absolute uncertainties for the signal and SM background, and the scale factors between
data and simulation, need to be taken into account. These scale factors correct for the differences in the lepton identification efficiency,
b-tagging efficiency, and pileup as described in Sections~\ref{sec:EventSamples} and \ref{sec:Systematics_experimentalMETT}.
The effect of the b-tagging efficiency is investigated by scaling the scale factors up and down in simulated events.
This is performed separately for the b-tagging efficiency scale factor and the mistagging rate scale factor.
Since triggers are not used in the simulation, scale factors are applied to account for the trigger
efficiencies when the simulation is compared to data. An additional uncertainty of 0.2\% accounts for the trigger efficiency correction
for the prediction in data.
The product of all scale factors differs from one by at most ten percent.

Model uncertainties are also taken into account. For the dominant \ttbar
background, the uncertainties for the inclusive cross section are calculated using the Monte Carlo for femtobarn processes (\MCFM~5.8)
\cite{MCFM}. The uncertainties associated with scales are determined by separately varying the factorization and matching scales
by a factor of 2 up and down. Including parton distribution function (PDF) uncertainties~\cite{Botje:2011sn}, we apply a total
uncertainty of 16\%.

The uncertainties for SM simulation in signal region $D$, shown in Table~\ref{tab:SM_sys}, are
needed for the comparison of data with the SM simulation (as shown in Section~\ref{sec:Result}), but are not used in
the limit determination with the scans.

\begin{table*}[tbh]
\topcaption{Systematic uncertainties in the signal region for the different selections for the SM simulation,
needed for the comparison with data (as in Table~\ref{tab:results}).
The signal regions corresponding to the number of required b tags are as defined in Table~\ref{tab:regions}.
All uncertainties are summed in quadrature. The variations in JES, JER, $\pt^\text{lepton}$, and unclustered energy are propagated to the $\ETslash$.
\label{tab:SM_sys}}
\scotchrule[l||c|c|c|c||c]
Variation & $\Delta N_D$ & $\Delta N_D$ & $\Delta N_D$ & $\Delta N_D$ & $\Delta N_D$ \\
& (0 b tags) & (1 b tag) & (2 b tags) & (${\ge}3$ b tags)& (${\ge}1$ b tag)\\
\hline %\hline
JES  & 17.8\%  & 16.7\%  & 19.2\%  & 17.3\%  & 17.5\% \\ %\hline
JER  & 17.1\%  & 4.8\%  & 6.2\%  & 5.4\%  & 5.3\% \\ %\hline
$\pt^\text{lepton}$   & 0.6\%  & 2.4\%  & 2.1\%  & 2.5\%  & 1.9\% \\ %\hline
Unclustered energy  & 0.1\%  & 0.9\%  & 1.1\%  & 0.5\%  & 1.0\% \\ %\hline
Pileup  & 2.7\%  & 2.2\%  & 0.8\%  & 1.1\%  & 1.6\% \\ %\hline
b-tagging scale factor  & 2.6\%  & 1.2\%  & 4.1\%  & 7.8\%  & 1.5\% \\ %\hline
Mistagging scale factor  & 2.0\%  & 0.8\%  & 1.3\%  & 5.9\%  & 1.3\% \\ %\hline
Model uncertainty  & 16.0\%  & 16.0\%  & 16.0\%  & 16.0\%  & 16.0\% \\ %\hline
Lepton trigger \& ID  & 3.0\%  & 3.0\%  & 3.0\%  & 3.0\%  & 3.0\% \\ %\hline
Luminosity uncertainty  & 2.2\%  & 2.2\%  & 2.2\%  & 2.2\%  & 2.2\% \\ \hline
\hline
Total uncertainty   & 30.0\%  & 26.7\%  & 26.5\%  & 24.4\%  & 24.8\% \\ \hline%\hline
Statistical uncertainty   & 11.0\%  & 8.3\%  & 8.6\%  & 3.4\%  & 5.6\% \\
\donescotchrule
\end{table*}

\section{Results}
\label{sec:Result}

The background estimation methods described in Sections
\ref{sec:SR_BG_METT} and \ref{sec:SR_BG_ABCD} are used to
predict the SM contribution to the signal regions.

\begin{figure}[htbp]
\begin{center}
\ifthenelse{\boolean{cms@external}}{
\includegraphics[width=.40\textwidth]{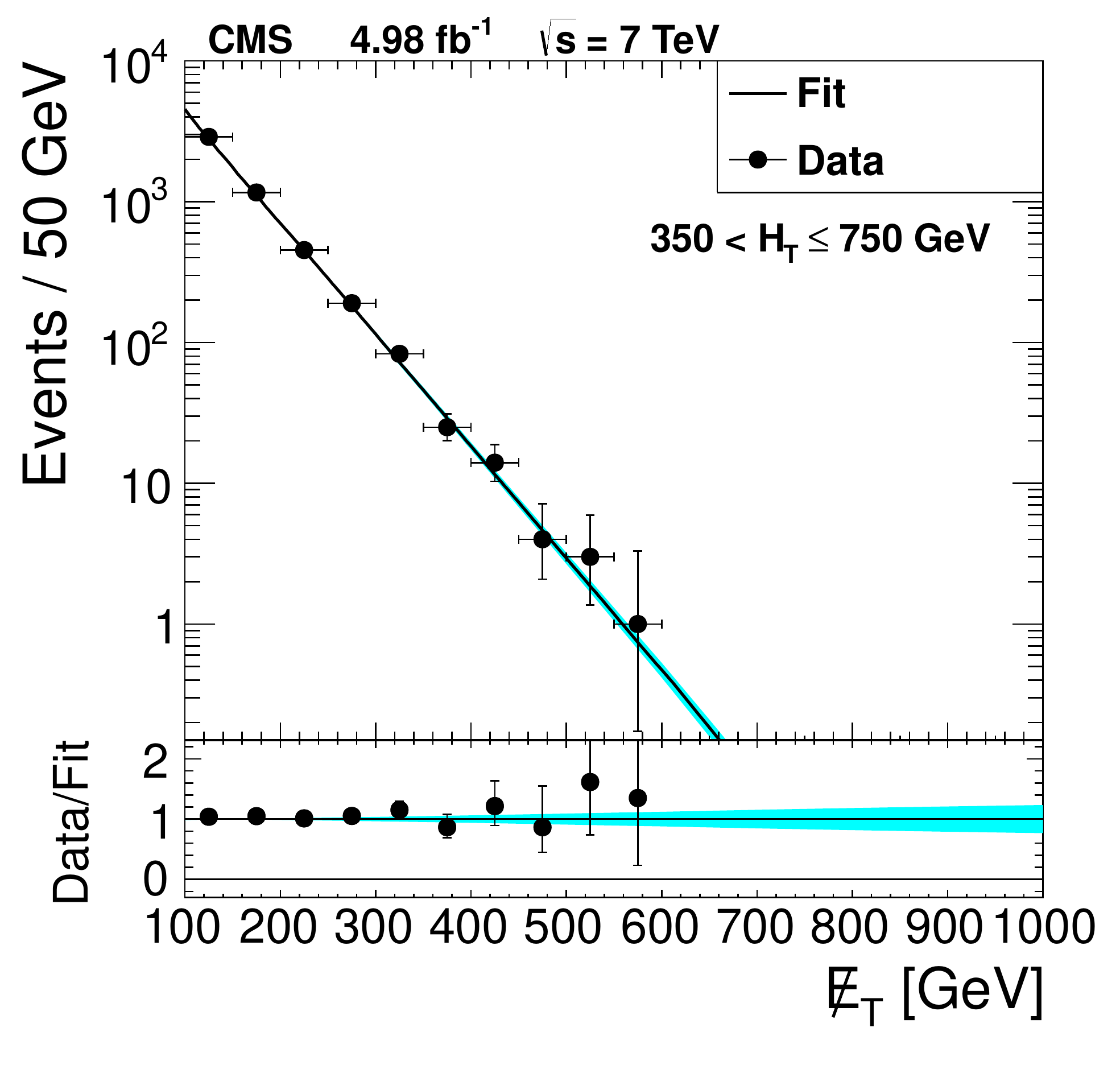}
\includegraphics[width=.40\textwidth]{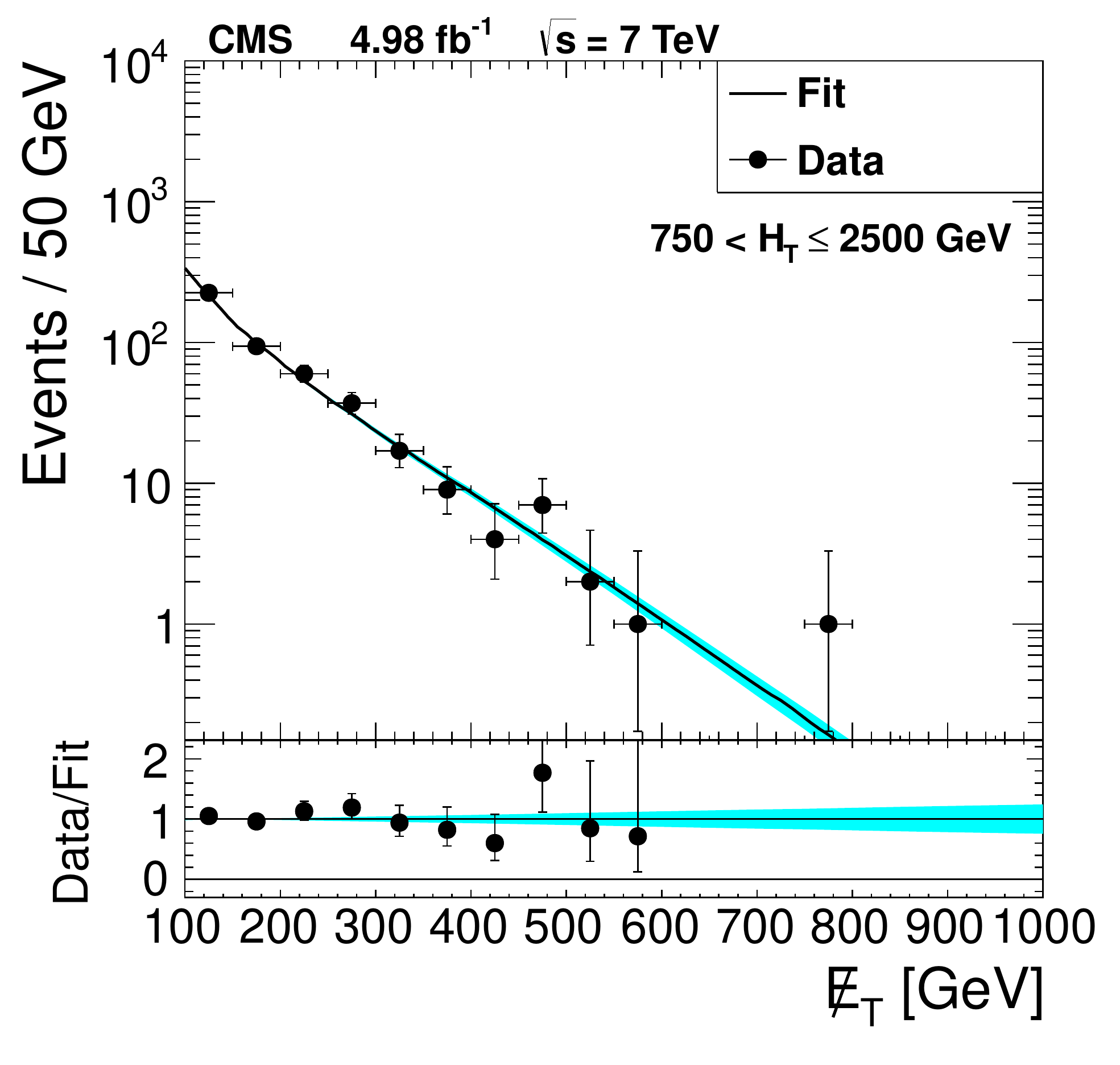}
\includegraphics[width=.40\textwidth]{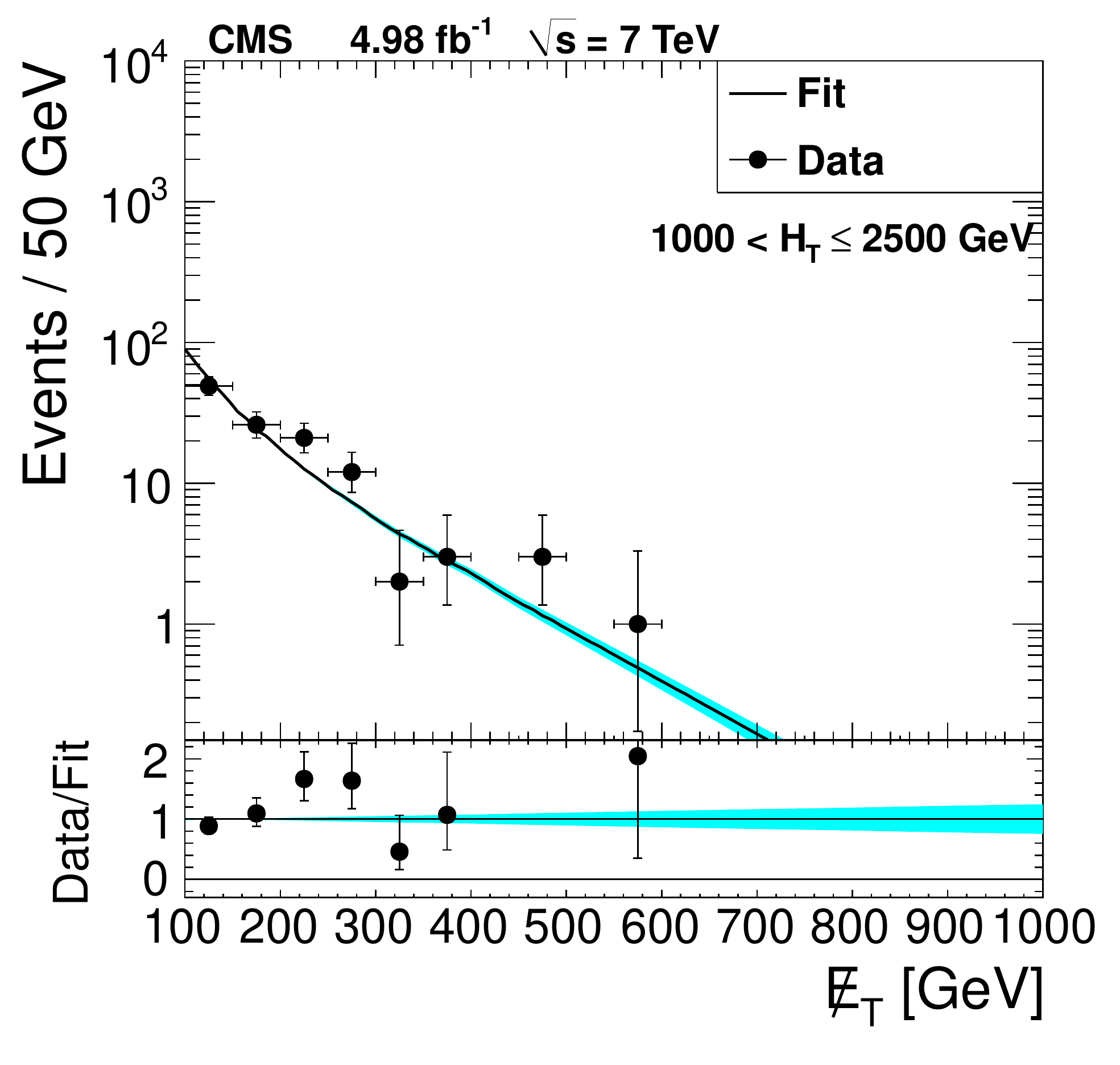}
}{
\includegraphics[width=.49\textwidth]{c_met_sumAllBins_ht-750_Data_lowht.pdf} \\
\includegraphics[width=.49\textwidth]{c_met_sumAllBins_ht-750_Data_highht.pdf}
\includegraphics[width=.49\textwidth]{c_met_sumAllBins_ht-1000_Data_highht.pdf}
}
\caption{
Distribution of \ETslash in the muon channel: data (points) and fit result of the template method (line) for (\cmsThreeLeft) $350 < \HT < 750\GeV$; data and prediction obtained from the fit for (\cmsThreeCenter) $750 < \HT < 2500\GeV$ and for (\cmsThreeRight) $1000 < \HT < 2500\GeV$.
The bands around the fit correspond to the statistical uncertainty on the parameter $\alpha$.
The systematic uncertainties have been calculated for $\ETslash >$ 250, 350 and 450 \GeV and range from 16 -- 32\% (24 -- 42\%) for $\HT >$ 750 (1000) \GeV as reported in Tab.~\ref{tab:final-prediction}.
The lower panels show the ratio between the fitted model and data.
}\label{fig:fitresult-data-metshape}
\end{center}
\end{figure}

A graphical representation of the \ETslash spectra estimated with the template method in a back\-ground-dominated region at low \HT and the two signal regions at high \HT are shown in Fig.~\ref{fig:fitresult-data-metshape}.
The fit provides a good description of the observed spectrum in the control region, and no excess is observed at high \HT.
The numerical results for different signal regions are summarized in Table~\ref{tab:final-prediction}, along with the observed event counts and the expectations for the two SUSY benchmark scenarios LM6 and LM8.
No events are observed above the common upper boundaries of the signal regions of $\HT < 2.5\TeV$ and $\ETslash < 2\TeV$.

\begin{table*}[bth]
\small
\topcaption{
Predicted and observed yields from the \ETslash-template method for the different signal regions.
The first uncertainties are statistical and the second systematic.
The expected yields and statistical uncertainties for the two benchmark points LM6 and LM8 are shown for comparison.
}\label{tab:final-prediction}
\scotchrule[l||r|rcrcr||rcr|rcr]
\hline & observed & predicted && stat.&&sys. & \multicolumn{3}{c}{LM6}& \multicolumn{3}{c}{LM8} \\
\hline & \multicolumn{12}{c}{$750 < \HT < 2500\GeV$, $250 < \ETslash < 2000\GeV$ } \\\hline
Total& 137 & 146\phantom{.00} & $\pm$ & 9\phantom{.00} & $\pm$ & 24\phantom{.00} & 42.2\phantom{0} & $\pm$ & 6.5\phantom{0} & 79.8\phantom{0} & $\pm$ & 8.9\phantom{0} \\
$0$ \cPqb\ tag& 97 & 99\phantom{.00} & $\pm$ & 8\phantom{.00} & $\pm$ & 18\phantom{.00} & 26.3\phantom{0} & $\pm$ & 5.1\phantom{0} & 21.9\phantom{0} & $\pm$ & 4.7\phantom{0} \\
$1$ \cPqb\ tag& 35 & 34.6\phantom{0} & $\pm$ & 2.8\phantom{0} & $\pm$ & 7.5\phantom{0} & 10.7\phantom{0} & $\pm$ & 3.3\phantom{0} & 29.0\phantom{0} & $\pm$ & 5.4\phantom{0} \\
${\ge}1$ \cPqb\ tag& 40 & 47\phantom{.00} & $\pm$ & 3\phantom{.00} & $\pm$ & 10\phantom{.00} & 16.0\phantom{0} & $\pm$ & 4.0\phantom{0} & 57.9\phantom{0} & $\pm$ & 7.6\phantom{0} \\
${\ge}2$ \cPqb\ tags& 5 & 12.3\phantom{0} & $\pm$ & 1.4\phantom{0} & $\pm$ & 2.7\phantom{0} & 5.2\phantom{0} & $\pm$ & 2.3\phantom{0} & 28.8\phantom{0} & $\pm$ & 5.4\phantom{0} \\
\hline & \multicolumn{12}{c}{$750 < \HT < 2500\GeV$, $350 < \ETslash < 2000\GeV$ } \\\hline
Total& 44 & 54\phantom{.00} & $\pm$ & 5\phantom{.00} & $\pm$ & 12\phantom{.00} & 30.7\phantom{0} & $\pm$ & 5.5\phantom{0} & 39.1\phantom{0} & $\pm$ & 6.3\phantom{0} \\
$0$ \cPqb\ tag& 32 & 38.7\phantom{0} & $\pm$ & 3.6\phantom{0} & $\pm$ & 9.5\phantom{0} & 19.9\phantom{0} & $\pm$ & 4.5\phantom{0} & 12.0\phantom{0} & $\pm$ & 3.5\phantom{0} \\
$1$ \cPqb\ tag& 11 & 11.5\phantom{0} & $\pm$ & 1.0\phantom{0} & $\pm$ & 3.5\phantom{0} & 7.5\phantom{0} & $\pm$ & 2.7\phantom{0} & 14.6\phantom{0} & $\pm$ & 3.8\phantom{0} \\
${\ge}1$ \cPqb\ tag& 12 & 14.8\phantom{0} & $\pm$ & 1.1\phantom{0} & $\pm$ & 4.5\phantom{0} & 10.9\phantom{0} & $\pm$ & 3.3\phantom{0} & 27.1\phantom{0} & $\pm$ & 5.2\phantom{0} \\
${\ge}2$ \cPqb\ tags& 1 & 3.3\phantom{0} & $\pm$ & 0.4\phantom{0} & $\pm$ & 1.0\phantom{0} & 3.3\phantom{0} & $\pm$ & 1.8\phantom{0} & 12.4\phantom{0} & $\pm$ & 3.5\phantom{0} \\
\hline & \multicolumn{12}{c}{$750 < \HT < 2500\GeV$, $450 < \ETslash < 2000\GeV$ } \\\hline
Total& 20 & 19.6\phantom{0} & $\pm$ & 2.1\phantom{0} & $\pm$ & 6.2\phantom{0} & 19.6\phantom{0} & $\pm$ & 4.4\phantom{0} & 15.8\phantom{0} & $\pm$ & 4.0\phantom{0} \\
$0$ \cPqb\ tag& 14 & 14.9\phantom{0} & $\pm$ & 1.7\phantom{0} & $\pm$ & 5.2\phantom{0} & 13.3\phantom{0} & $\pm$ & 3.7\phantom{0} & 5.7\phantom{0} & $\pm$ & 2.4\phantom{0} \\
$1$ \cPqb\ tag& 5 & 3.8\phantom{0} & $\pm$ & 0.4\phantom{0} & $\pm$ & 1.5\phantom{0} & 4.5\phantom{0} & $\pm$ & 2.1\phantom{0} & 5.6\phantom{0} & $\pm$ & 2.4\phantom{0} \\
${\ge}1$ \cPqb\ tag& 6 & 4.7\phantom{0} & $\pm$ & 0.4\phantom{0} & $\pm$ & 1.8\phantom{0} & 6.3\phantom{0} & $\pm$ & 2.5\phantom{0} & 10.1\phantom{0} & $\pm$ & 3.2\phantom{0} \\
${\ge}2$ \cPqb\ tags& 1 &      0.9\phantom{0}             & $\pm$ &    0.1\phantom{0}            & $\pm$ &    0.3\phantom{0}            &  1.8\phantom{0}           & $\pm$ &    1.3\phantom{0}            &     4.5\phantom{0}           & $\pm$ &    2.1\phantom{0} \\
\hline & \multicolumn{12}{c}{$1000 < \HT < 2500\GeV$, $250 < \ETslash < 2000\GeV$ } \\\hline
Total& 36 & 37.5\phantom{0} & $\pm$ & 3.7\phantom{0} & $\pm$ & 8.9\phantom{0} & 18.1\phantom{0} & $\pm$ & 4.3\phantom{0} & 31.1\phantom{0} & $\pm$ & 5.6\phantom{0} \\
$0$ \cPqb\ tag& 30 & 27.0\phantom{0} & $\pm$ & 3.2\phantom{0} & $\pm$ & 7.0\phantom{0} & 10.9\phantom{0} & $\pm$ & 3.3\phantom{0} & 7.9\phantom{0} & $\pm$ & 2.8\phantom{0} \\
$1$ \cPqb\ tag& 5 & 7.5\phantom{0} & $\pm$ & 1.2\phantom{0} & $\pm$ & 2.6\phantom{0} & 4.8\phantom{0} & $\pm$ & 2.2\phantom{0} & 11.5\phantom{0} & $\pm$ & 3.4\phantom{0} \\
${\ge}1$ \cPqb\ tag& 6 & 10.5\phantom{0} & $\pm$ & 1.3\phantom{0} & $\pm$ & 3.6\phantom{0} & 7.2\phantom{0} & $\pm$ & 2.7\phantom{0} & 23.2\phantom{0} & $\pm$ & 4.8\phantom{0} \\
${\ge}2$ \cPqb\ tags& 1 & 3.0\phantom{0} & $\pm$ & 0.6\phantom{0} & $\pm$ & 1.0\phantom{0} & 2.3\phantom{0} & $\pm$ & 1.5\phantom{0} & 11.8\phantom{0} & $\pm$ & 3.4\phantom{0} \\
\hline & \multicolumn{12}{c}{$1000 < \HT < 2500\GeV$, $350 < \ETslash < 2000\GeV$ } \\\hline
Total& 13 & 15.5\phantom{0} & $\pm$ & 1.7\phantom{0} & $\pm$ & 4.9\phantom{0} & 13.0\phantom{0} & $\pm$ & 3.6\phantom{0} & 15.6\phantom{0} & $\pm$ & 4.0\phantom{0} \\
$0$ \cPqb\ tag& 11 & 11.7\phantom{0} & $\pm$ & 1.6\phantom{0} & $\pm$ & 4.2\phantom{0} & 8.1\phantom{0} & $\pm$ & 2.8\phantom{0} & 4.2\phantom{0} & $\pm$ & 2.1\phantom{0} \\
$1$ \cPqb\ tag& 2 & 2.9\phantom{0} & $\pm$ & 0.5\phantom{0} & $\pm$ & 1.4\phantom{0} & 3.4\phantom{0} & $\pm$ & 1.9\phantom{0} & 5.8\phantom{0} & $\pm$ & 2.4\phantom{0} \\
${\ge}1$ \cPqb\ tag& 2 & 3.8\phantom{0} & $\pm$ & 0.5\phantom{0} & $\pm$ & 1.8\phantom{0} & 4.9\phantom{0} & $\pm$ & 2.2\phantom{0} & 11.4\phantom{0} & $\pm$ & 3.4\phantom{0} \\
${\ge}2$ \cPqb\ tags& 0 &      1.0\phantom{0}             & $\pm$ &   0.2\phantom{0}            & $\pm$ &   0.5\phantom{0}              &   1.5\phantom{0} & $\pm$           &    1.2\phantom{0}            &     5.6\phantom{0} & $\pm$            &    2.4\phantom{0}  \\
\hline & \multicolumn{12}{c}{$1000 < \HT < 2500\GeV$, $450 < \ETslash < 2000\GeV$ } \\\hline
Total& 7 & 6.6\phantom{0} & $\pm$ & 0.9\phantom{0} & $\pm$ & 2.8\phantom{0} & 8.5\phantom{0} & $\pm$ & 2.9\phantom{0} & 7.0\phantom{0} & $\pm$ & 2.6\phantom{0} \\
$0$ \cPqb\ tag& 6 & 5.2\phantom{0} & $\pm$ & 0.8\phantom{0} & $\pm$ & 2.3\phantom{0} & 5.5\phantom{0} & $\pm$ & 2.3\phantom{0} & 2.2\phantom{0} & $\pm$ & 1.5\phantom{0} \\
$1$ \cPqb\ tag& 1 &       1.1\phantom{0}  & $\pm$ &   0.2\phantom{0}  & $\pm$ &   0.7\phantom{0} &   2.1\phantom{0} & $\pm$ &    1.5\phantom{0}  &     2.4\phantom{0} & $\pm$ &    1.6\phantom{0} \\
${\ge}1$ \cPqb\ tag& 1 &  1.4\phantom{0}  & $\pm$ &   0.2\phantom{0}  & $\pm$ &   0.9\phantom{0} &   3.0\phantom{0} & $\pm$ &    1.7\phantom{0}  &     4.8\phantom{0} & $\pm$ &    2.2\phantom{0} \\
${\ge}2$ \cPqb\ tags& 0 &       0.3\phantom{0}  & $\pm$ &   0.1\phantom{0}  & $\pm$ &   0.2\phantom{0} &   0.8\phantom{0} & $\pm$ &    0.9\phantom{0}  &     2.4\phantom{0} & $\pm$ &    1.5\phantom{0} \\
\donescotchrule
\end{table*}

For the factorization method, the number of events in the signal region $N_D$ and the predicted
value $\hat{N}_D$ are summarized in Table~\ref{tab:results}, which additionally includes expectations
for the SM and for the SM with contributions of
the LM6 and LM8 SUSY scenarios added.
The measured number of events $N_D$ and the predicted value $\hat{N}_D$ are in agreement and no
excess is observed. The reconstructed number of events in region $D$ and the predicted value $\hat{N}_D$ are in agreement
also for the SM simulation, showing the validity of the factorization
ansatz for the background estimation. For the comparison of data and simulation,
several scale factors are taken into account, as described in Section~\ref{sec:Systematics_ABCD}. The uncertainty on the number of events $N_D$ for the SM prediction from simulation is larger than that on the prediction $\hat{N}_D$
from data, showing the advantage of this background estimation method.

\begin{table*}[htb]
\topcaption{Number of reconstructed ($N_D$) and predicted ($\hat{N}_D$) events in the signal region for the
factorization method for the SM, two possible signal scenarios (LM6, LM8), and data. The first uncertainties are statistical and the second systematic.
The systematic uncertainty on $\hat{N}_D$ in data is equal to the uncertainty on $\kappa$.
The systematic uncertainty in simulation includes the uncertainty on the absolute rate of simulated events, as discussed in the text.
The exclusive 0 b-tag selection is shown for comparison as well.}\label{tab:results}
\scotchrule[l||l|c|c]
Signal region & Sample & $N_D$ &  $\hat{N}_D$ \\
\hline \hline
\multirow{4}{*}{\parbox{10em}{ 0 b-tags \\ $\HT > 600$\GeV \\ $Y_\mathrm{MET}>6.5$~$\sqrt{\GeVns{}}$ \rule{0pt}{1.1em} }}

& $\Sigma$ SM     & 182 $\pm$ 22 $\pm$ 55 & 186 $\pm$ 19 $\pm$ 40 \\
& $\Sigma$ SM+LM6 & 221 $\pm$ 22 $\pm$ 59 & 191 $\pm$ 19 $\pm$ 40 \\
& $\Sigma$ SM+LM8 & 218 $\pm$ 24 $\pm$ 61 & 194 $\pm$ 20 $\pm$ 41 \\
\cline{2-4}
& Data            & 155        & 162 $\pm$ 11 $\pm$ 18 \\
\hline

\multirow{4}{*}{\parbox{10em}{ 1 b-tag\\ $\HT > 800$\GeV \\ $Y_\mathrm{MET}>5.5$~$\sqrt{\GeVns{}}$ \rule{0pt}{1.1em} }}
& $\Sigma$ SM     &  74 $\pm$  5 $\pm$ 18 &  74 $\pm$  4 $\pm$ 14 \\
& $\Sigma$ SM+LM6 &  95 $\pm$  5 $\pm$ 21 &  77 $\pm$  4 $\pm$ 14 \\
& $\Sigma$ SM+LM8 & 132 $\pm$  6 $\pm$ 29 &  90 $\pm$  5 $\pm$ 16 \\
\cline{2-4}
& Data            &  51                         &  53.9 $\pm$  6.3 $\pm$  5.9 \\
\hline

\multirow{4}{*}{\parbox{10em}{ 2 b-tags \\ $\HT > 800$\GeV \\ $Y_\mathrm{MET}>5.5$~$\sqrt{\GeVns{}}$ \rule{0pt}{1.1em} }}
& $\Sigma$ SM     &  50 $\pm$  3 $\pm$ 13 &  47.5 $\pm$  2.1 $\pm$  8.1 \\
& $\Sigma$ SM+LM6 &  62 $\pm$  3 $\pm$ 15 &  49.0 $\pm$  2.2 $\pm$  8.2 \\
& $\Sigma$ SM+LM8 & 103 $\pm$  5 $\pm$ 24 &  62.7 $\pm$  2.7 $\pm$  9.7 \\
\cline{2-4}
& Data            &  27                         &  36.0 $\pm$  5.1 $\pm$  4.0 \\
\hline

\multirow{4}{*}{\parbox{10em}{ $\ge 3$ b-tags \\ $\HT > 600$\GeV \\ $Y_\mathrm{MET}>6.5$~$\sqrt{\GeVns{}}$ \rule{0pt}{1.1em} }}
& $\Sigma$ SM     & 22.6 $\pm$  1.1 $\pm$  6.0 &  21.3 $\pm$  0.9 $\pm$  4.0 \\
& $\Sigma$ SM+LM6 & 27.1 $\pm$  1.1 $\pm$  6.6 &  21.9 $\pm$  0.9 $\pm$  4.1 \\
& $\Sigma$ SM+LM8 & 66 $\pm$  4 $\pm$ 15 &  34.3 $\pm$  1.8 $\pm$  4.8 \\
\cline{2-4}
& Data            & 10                         &  13.8 $\pm$  3.2 $\pm$  1.5 \\
\hline

\multirow{4}{*}{\parbox{10em}{ $\ge 1$ b-tag\\ $\HT > 800$\GeV \\ $Y_\mathrm{MET}>5.5$~$\sqrt{\GeVns{}}$ \rule{0pt}{1.1em} }}
& $\Sigma$ SM     & 136 $\pm$  6 $\pm$ 34 & 134 $\pm$  5 $\pm$ 24 \\
& $\Sigma$ SM+LM6 & 172 $\pm$  6 $\pm$ 39 & 139 $\pm$  5 $\pm$ 24 \\
& $\Sigma$ SM+LM8 & 280 $\pm$  8 $\pm$ 63 & 177 $\pm$  6 $\pm$ 28 \\
\cline{2-4}
& Data            &  84                         &  98 $\pm$  8 $\pm$ 11 \\

\donescotchrule
\end{table*}

\section{Interpretation} \label{sec:Interpretation}

Using the results presented in Section~\ref{sec:Result}, limits are set on the parameters of
several supersymmetric models, including the CMSSM and
the simplified model described in Section~\ref{sec:EventSamples}.

Limits are set using the \CLs method~\cite{Read:2002cls, Junk:1999kv} with
a test statistic given by a profile likelihood ratio.
The likelihood function includes a Poisson distribution describing the number of observed events in the signal region.
Its mean value is $B + \mu S$, where $B$ is the predicted background, $S$ the expected signal yield at the nominal cross section of the model under study, and $\mu$ the signal strength parameter.

For the \ETslash template method, $B = B_\mathrm{N} r / (1+\mu c)$, where $B_\mathrm{N}$ is the background in the normalization region, $r$ the ratio of the background in signal and normalization regions, determined by the \ETslash model, and $c$ is the relative bias in the background estimation due to signal contamination.
The effect of signal contamination is determined by repeating the background estimation on simulated samples com\-bining SM processes and a signal at the nominal cross section.
The nuisance parameter $B_\mathrm{N}$ is constrained by a second Poisson distribution with mean $B_\mathrm{N}$, describing the number of observed events in the normalization region.
For the factorization method, $B = \kappa B_\mathrm{B} B_\mathrm{C} / B_\mathrm{A}$.
The nuisance parameters $B_i$ describing the estimated background in the three control regions A, B, and C are constrained by three additional Poisson distributions with mean values $B_i + \mu \alpha_i S$, where $i$ is the index of a control region.
The second term describes the expected contribution of the signal to the control region and ensures a correct estimate in the presence of signal contamination.
The full likelihood function contains additional log-normal terms describing the nuisance parameters affecting the expected signal yields and the parameters $r$ and $\kappa$ for the \ETslash template and the factorization method, respectively, corresponding to the different sources of systematic uncertainties.

The expected signal yields and systematic uncertainties are evaluated for every signal point in the parameter planes of the two models considered.
Sources of experimental uncertainties on the signal selection include the jet energy and \ETslash scales, \cPqb-tagging efficiencies, and mistagging rates.
These uncertainties are treated as fully correlated with the corresponding variations in the background estimate.
Smaller contributions to the signal uncertainty are due to the lepton and trigger selection efficiencies and to the measurement of the luminosity (2.2\%).
In the likelihood function used for the factorization method, the correlation of uncertainties between the four regions is taken into account.

\subsection{CMSSM}

Within the CMSSM limits are set in the $m_{1/2}$ vs.\ $m_{0}$ plane with parameters $\tan \beta = 10$, $A_0 = 0$\GeV, and $\mu > 0$.
The acceptance and efficiency factors $\epsilon_i\mathcal{A}_i$ are calculated in a scan over the parameters $m_0$ and $m_{1/2}$.
This is done with leading order (LO) simulation, combined with next-to-leading order (NLO) and
next-to-leading log (NLL) $K$-factors~\cite{Beenakker:1996ch, Kulesza:2008jb, Kulesza:2009kq, Beenakker:2009ha, Beenakker:2011fu} for each SUSY subprocess separately.
The experimental uncertainties on the signal selection efficiency are dominated by the jet and \ETslash energy scales.
In the relevant region of the parameter plane, these variations are smaller than 20\% for both methods.
The contributions due to the lepton and trigger selection are about $5\%$.

For the \ETslash template method, the CMSSM limits are set in a multichannel approach using the \zeroTag, \oneTag, and \twoTag bins, while for the factorization method at least one b tag is required.
In the multichannel approach, the statistical uncertainties on the background estimation due to fluctuations in the normalization regions are treated as uncorrelated.
Correlations between \cPqb-jet multiplicity bins in the \ETslash template method are evaluated for the uncertainties related to the \ETslash shape parameters.
Variations in the \cPqb-jet identification efficiencies also lead to correlation between different bins and between signal yields and background predictions.
All other systematic effects are treated as fully correlated.

\begin{figure}[htbp]
 \begin{center}
  \includegraphics[width=0.5\textwidth]{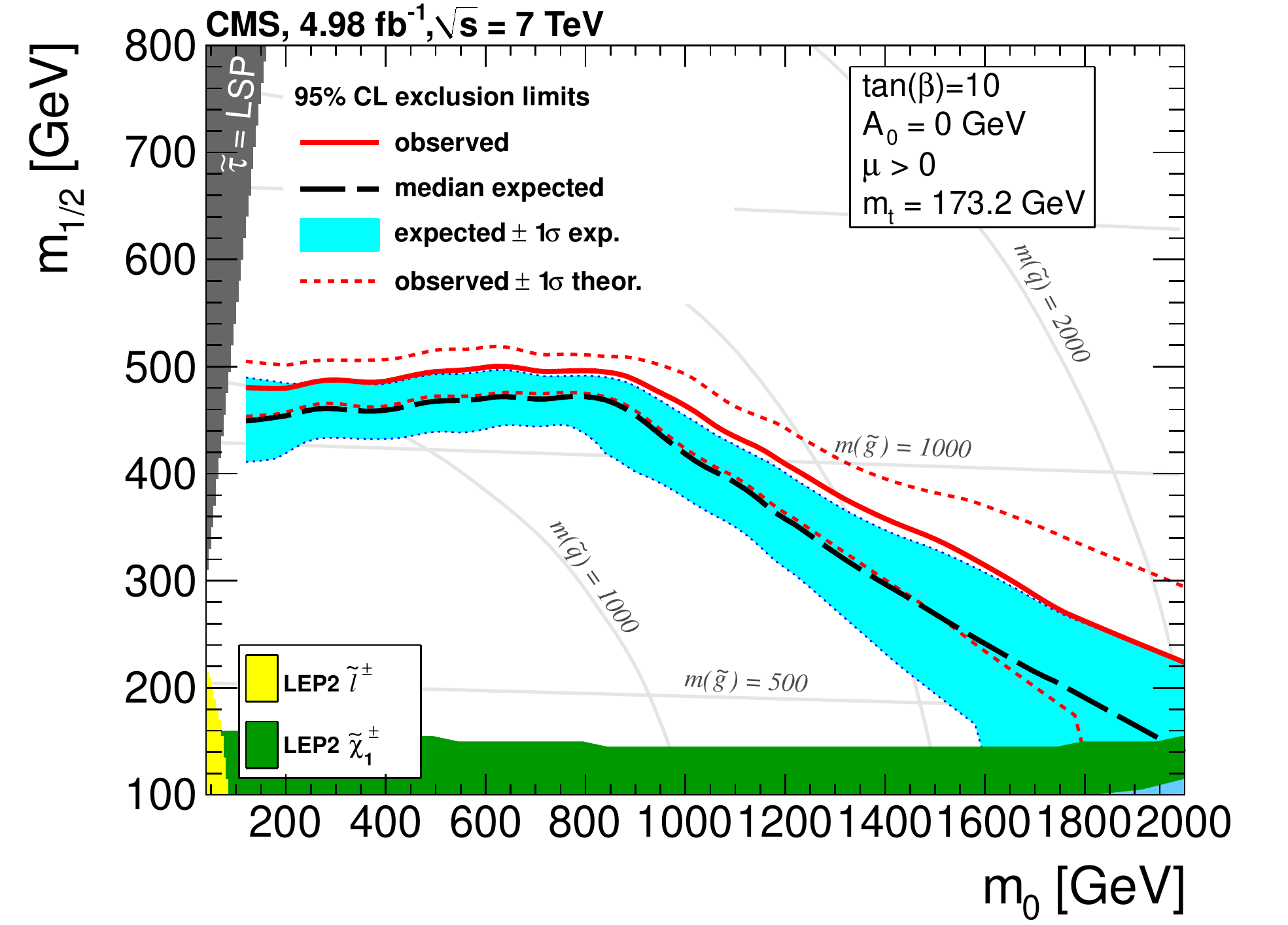}
  \includegraphics[width=0.5\textwidth]{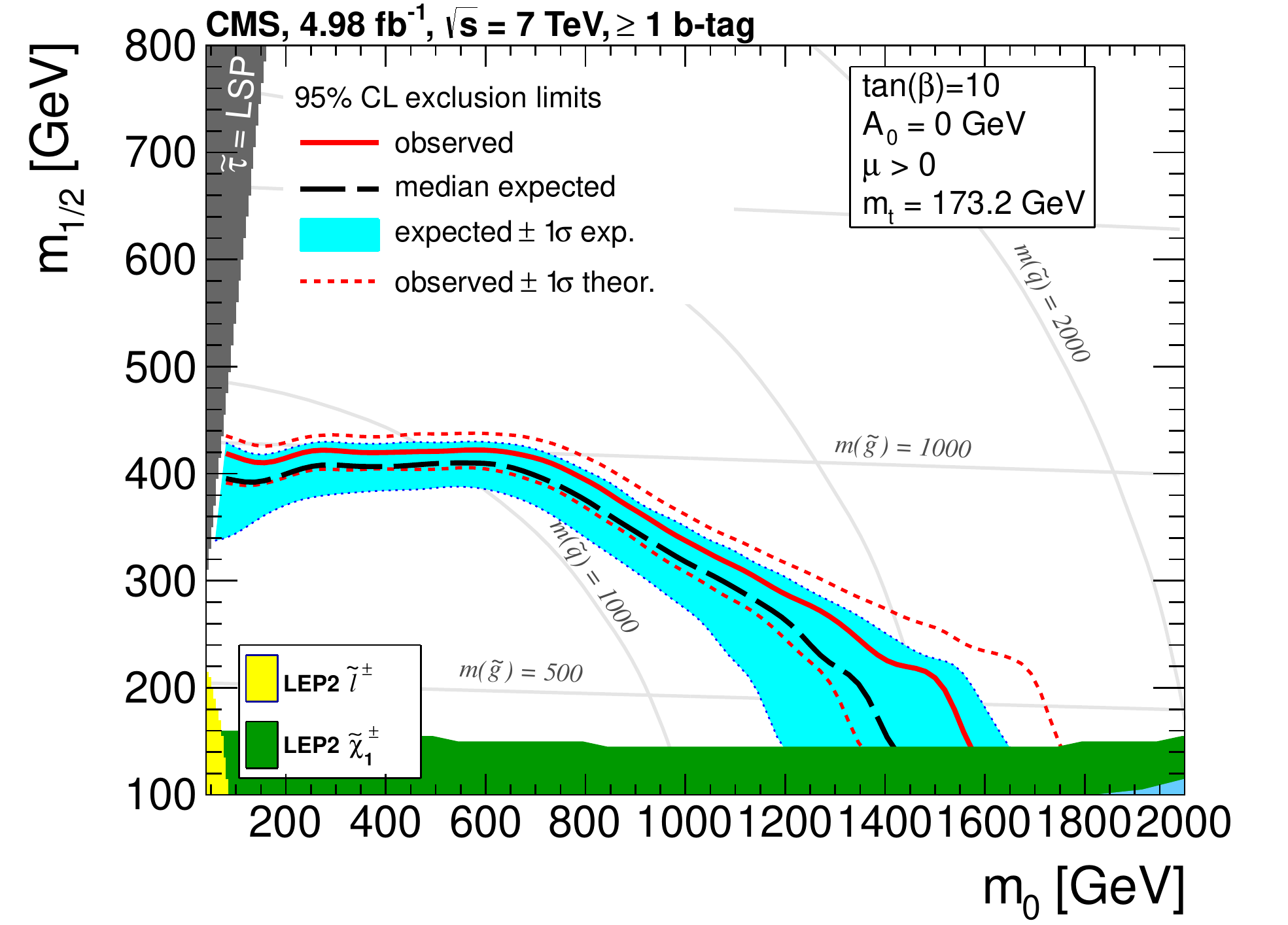}
 \end{center}
 \caption{The 95\% CL limit using the \CLs technique for the
   CMSSM model with $\tan\beta=10$, $A_0=0\GeV$, and $\mu>0$ (\cmsLeft)
for the \ETslash template method using the multichannel approach and
(\cmsRight) for the factorization method requiring at least one b tag.
The solid red line corresponds to the median expected limit, including all experimental uncertainties.
The area below the solid red line (observed limit) is excluded, with the thin red dashed lines showing the effect of a variation
of the signal yields due to theoretical uncertainties. The thick black dashed line shows the expected limit. It is surrounded by a shaded area
representing the experimental uncertainties.
 }\label{f.cmssmt10limitplot}
\end{figure}

The 95\% confidence level (CL) limit using the \CLs technique is presented in Fig.~\ref{f.cmssmt10limitplot}, where the region below the black curves is excluded.
The regions in \HT and \ETslash with the highest sensitivity are used: $\HT > 1000\GeV$ and $\ETslash > 250\GeV$ for the \ETslash template method, and $\HT > 800\GeV$, $Y_\mathrm{MET} > 5.5\,\sqrt{\GeVns{}}$, and ${\ge}1$ b tag for the factorization method.
Theoretical uncertainties on cross sections, arising from scale and  PDF uncertainties, are illustrated by bands of the expected and observed limits with these uncertainties added or subtracted \cite{Kramer:2012bx}.
The \ETslash template method with the simultaneous use of three \cPqb-jet multiplicity bins provides the best expected limit.

\subsection{Simplified model interpretation}

\begin{figure}[htb]
 \begin{center}
   \includegraphics[angle=0,width=0.49\textwidth]{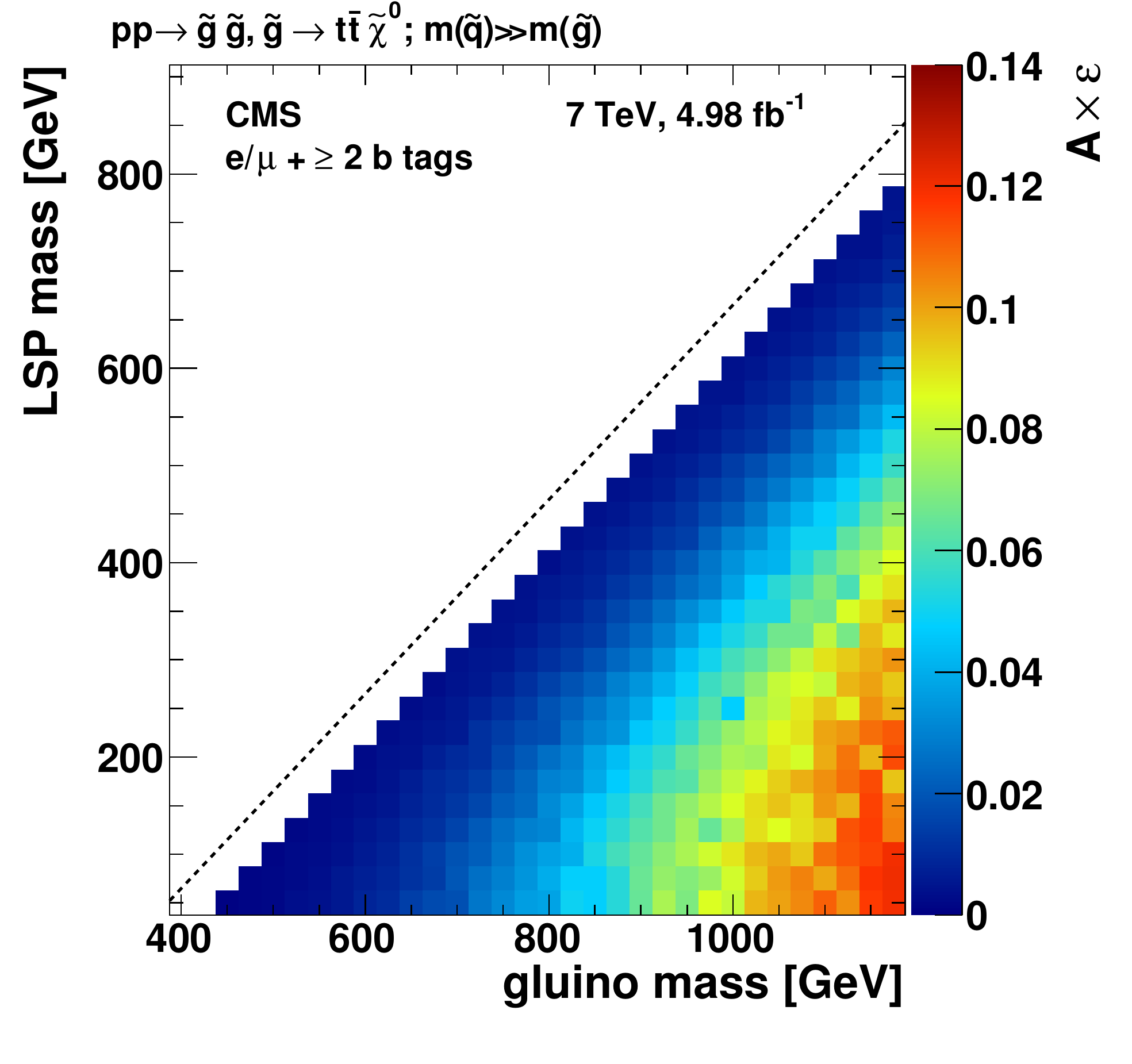}
   \includegraphics[angle=0,width=0.49\textwidth]{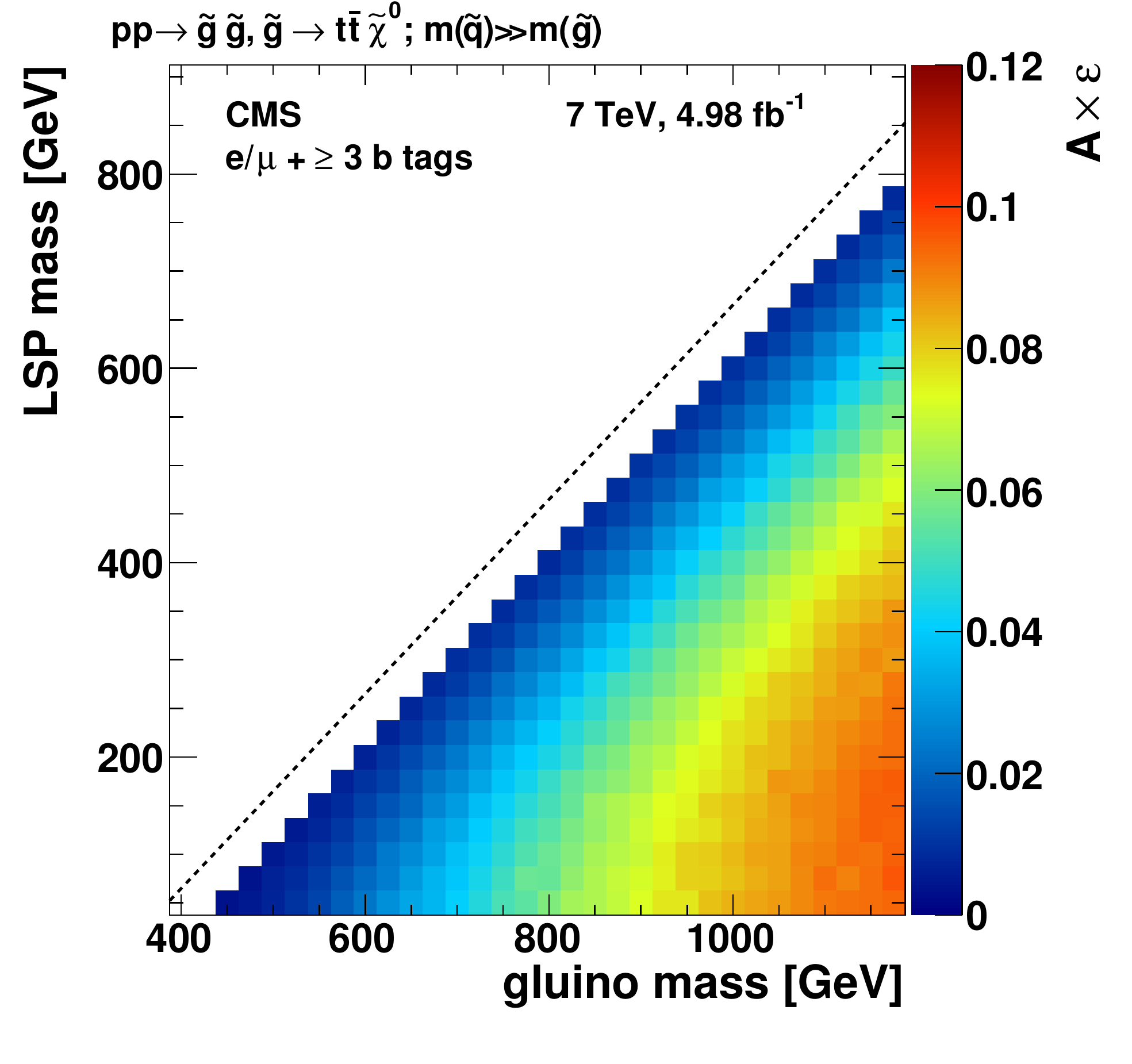}
 \end{center}
 \caption{
  Acceptance times efficiency for the simplified model shown in Fig.~\ref{fig:T1tttt} for (\cmsLeft) the \ETslash template method, where at least two
  b tags are required, and (\cmsRight) the factorization method with three or more b tags. The diagonal dashed line marks the lower kinematical limit of the LSP mass.
   }\label{fig:ea}
\end{figure}

In simplified models a limited set of hypothetical particles is introduced to produce a given
topological signature \cite{ArkaniHamed:2007fw,Alves:2011wf,Alwall:2008ag}. The final state of the simplified model studied here contains a lepton and
\cPqb\ jets as described in Section~\ref{sec:EventSamples}.
The model has no intermediate mass state, so it contains only two free parameters: the mass of the LSP and the mass of the gluino.
The signal cross sections are calculated up to NLO + NLL accuracy~\cite{Beenakker:1996ch, Kulesza:2008jb, Kulesza:2009kq, Beenakker:2009ha, Beenakker:2011fu, Kramer:2012bx}.
For each point in
the parameter plane, the acceptance times efficiency and a cross section upper-limit is calculated. The systematic uncertainties are, as in the CMSSM case,
determined for each point.
The acceptance times the efficiency is presented in Fig.~\ref{fig:ea} for both background estimation methods.

For the \ETslash template method, the best expected limits for
this model are obtained in the \twoTag bin.
Cross section limits at 95\% CL are calculated using the statistical
framework described above.
The signal region defined by the lower boundaries $\HT > 750\GeV$ and $\ETslash > 250\GeV$ is used. This choice results in high signal efficiencies also for low gluino
masses and small mass differences between the gluino and the LSP.
The limit with the factorization method is set requiring ${\ge}3$ b tags. In this case the signal
region is defined as $\HT > 600\GeV$ and $Y_\mathrm{MET} > 6.5\,\sqrt{\GeVns{}}$.

\begin{figure}[htb]
 \begin{center}
  \includegraphics[width=0.49\textwidth]{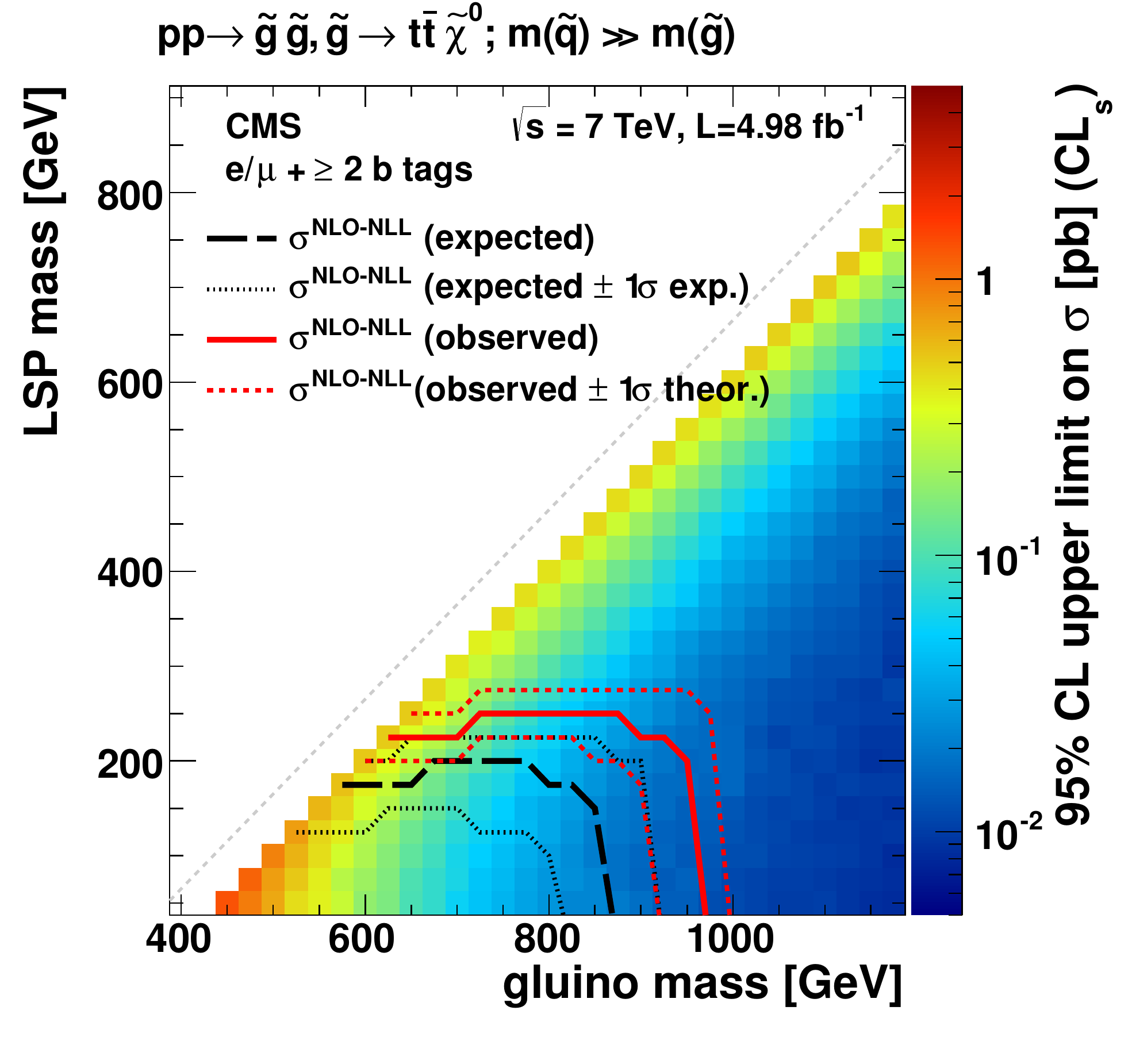}
  \includegraphics[width=0.49\textwidth]{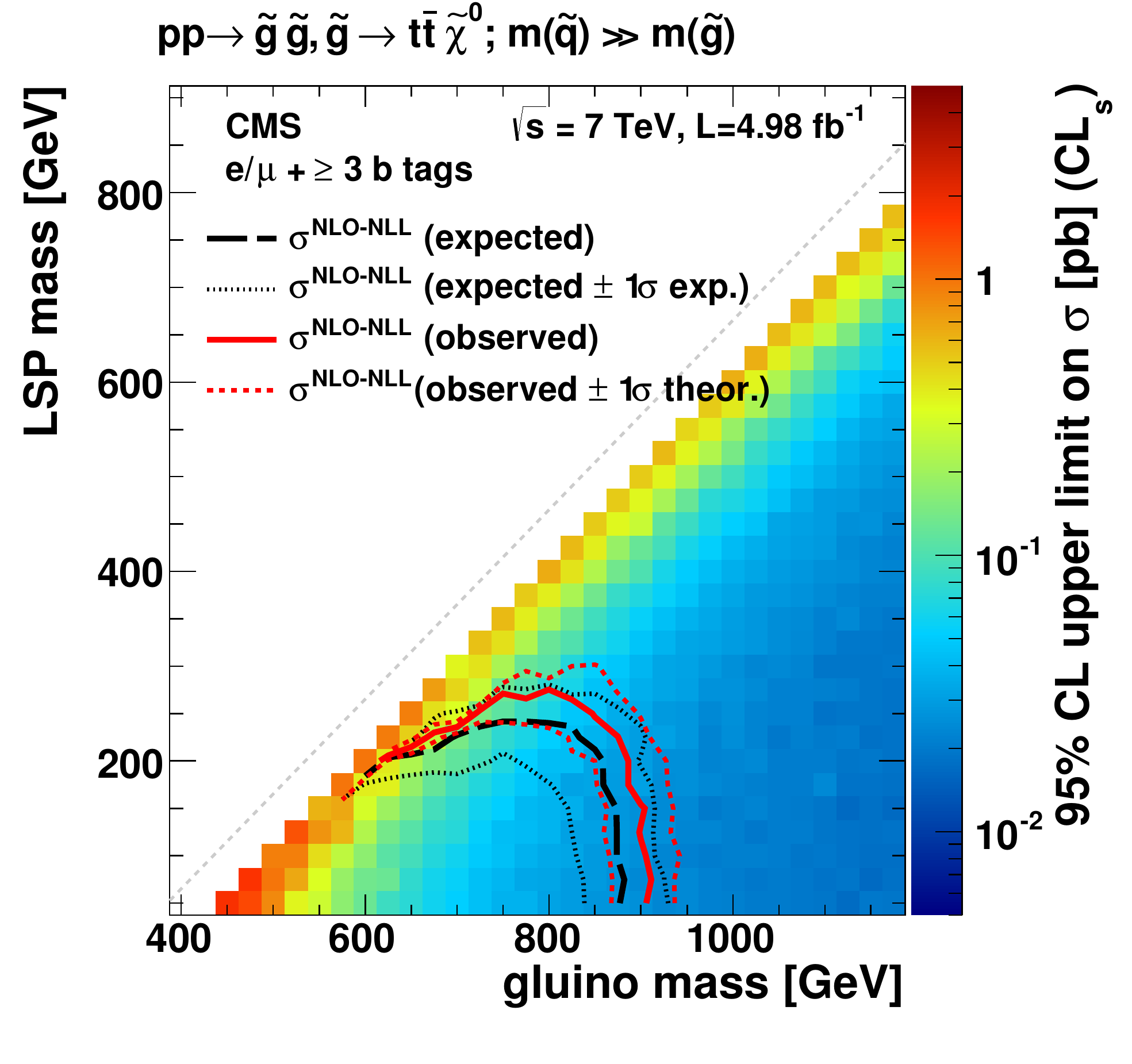}
 \end{center}
 \caption{
  The 95\% CL upper limit on the cross section using the \CLs technique for the simplified model shown in Fig.~\ref{fig:T1tttt} for (\cmsLeft) the \ETslash template method, where at least two
  b tags are required, and (\cmsRight) the factorization method with three or more b tags. The area below the thick solid red line
   is excluded. The thick dashed black line represents the expected limit. The diagonal dashed line marks the lower kinematical limit of the LSP mass.
  }\label{fig:t1ttttLimit}
\end{figure}

The effect of signal contamination on the background estimation is found to
be higher than in the CMSSM case, with values up to 30\%.
This bias is taken into account in the calculation of the limits, which are shown in Fig.~\ref{fig:t1ttttLimit}.

The analyses have also been tested on a simplified model describing direct stop pair production. Despite a higher acceptance times efficiency for this model,
no limits can be obtained due to the low cross section of this process.

\section{Summary}\label{sec:Conclusion}

A sample of events with a single electron or muon, multiple energetic jets, including identified \cPqb\ jets, and significant missing transverse energy,
has been used to perform a search for new physics motivated by $R$-parity conserving supersymmetric models.
The study is based on a data sample of proton-proton collision data recorded at $\sqrt{s}=7\TeV$ with the CMS detector, corresponding to an integrated
luminosity of 4.98\fbinv. The dominant standard model backgrounds are due to \ttbar and \Wjets production.

Background contributions to different signal regions have been estimated from data with two complementary approaches.
The first approach uses data in a control region at low \HT to determine templates of the \ETslash spectra for each of the main background components.
Fits are performed simultaneously for three subsamples with 0, 1, and $\ge2$ identified \cPqb\ jets to determine the templates.
Based on the templates and the observed number of events in a normalization region at low \ETslash, predictions are made for several signal regions at high \HT and \ETslash.
The second approach uses the low correlation between \HT and $Y_\mathrm{MET} = \ETslash / \sqrt{\HT}$.
The standard model background in signal regions at high values of \HT and $Y_\mathrm{MET}$ is estimated based on the observed yields in three control regions. The two background estimation
methods are complementary, as they have only small overlap in their control and signal
regions, both in the standard model and in the signal scenarios.

No excess has been observed, and the results have been used to set 95\% CL exclusion limits for several models.
In the context of the constrained minimal supersymmetric extension of the standard model with parameters $\tan\beta = 10$, $A_0 = 0\GeV$, and $\mu > 0$, the template method with the simultaneous
use of the 0, 1, and ${\ge}2$ \cPqb-jet bins shows the highest sensitivity.
Values of $m_{1/2}$ below about 450\GeV are excluded for $m_0$ in the range of about 200\GeV to about 800\GeV.

Limits have also been set in the parameter plane of the gluino and LSP masses of a simplified model that features four top quarks in the final state.
Due to the high number of \cPqb\ quarks in the final state, the factorization method, which provides a background estimate for events with at least
three identified \cPqb\ jets, has the highest sensitivity.
Using the SUSY production cross section as a reference, the exclusion reaches to gluino masses of about 870\GeV.
At a gluino mass of 750\GeV, LSP masses below 240\GeV are excluded.
This is the first CMS analysis of this scenario in the final state with a single lepton and b-tagged jets. A similar mass range is excluded by other
CMS analyses based on 2011 data \cite{Chatrchyan:1474300, Chatrchyan:2012jx, Chatrchyan:2012sa}.
Direct stop pair production can not yet be excluded with this analysis due to its low cross section.

\section*{Acknowledgements}
\hyphenation{Bundes-ministerium Forschungs-gemeinschaft Forschungs-zentren} We congratulate our colleagues in the CERN accelerator departments for the excellent performance of the LHC and thank the technical and administrative staffs at CERN and at other CMS institutes for their contributions to the success of the CMS effort. In addition, we gratefully acknowledge the computing centres and personnel of the Worldwide LHC Computing Grid for delivering so effectively the computing infrastructure essential to our analyses. Finally, we acknowledge the enduring support for the construction and operation of the LHC and the CMS detector provided by the following funding agencies: the Austrian Federal Ministry of Science and Research; the Belgian Fonds de la Recherche Scientifique, and Fonds voor Wetenschappelijk Onderzoek; the Brazilian Funding Agencies (CNPq, CAPES, FAPERJ, and FAPESP); the Bulgarian Ministry of Education, Youth and Science; CERN; the Chinese Academy of Sciences, Ministry of Science and Technology, and National Natural Science Foundation of China; the Colombian Funding Agency (COLCIENCIAS); the Croatian Ministry of Science, Education and Sport; the Research Promotion Foundation, Cyprus; the Ministry of Education and Research, Recurrent financing contract SF0690030s09 and European Regional Development Fund, Estonia; the Academy of Finland, Finnish Ministry of Education and Culture, and Helsinki Institute of Physics; the Institut National de Physique Nucl\'eaire et de Physique des Particules~/~CNRS, and Commissariat \`a l'\'Energie Atomique et aux \'Energies Alternatives~/~CEA, France; the Bundesministerium f\"ur Bildung und Forschung, Deutsche Forschungsgemeinschaft, and Helmholtz-Gemeinschaft Deutscher Forschungszentren, Germany; the General Secretariat for Research and Technology, Greece; the National Scientific Research Foundation, and National Office for Research and Technology, Hungary; the Department of Atomic Energy and the Department of Science and Technology, India; the Institute for Studies in Theoretical Physics and Mathematics, Iran; the Science Foundation, Ireland; the Istituto Nazionale di Fisica Nucleare, Italy; the Korean Ministry of Education, Science and Technology and the World Class University program of NRF, Korea; the Lithuanian Academy of Sciences; the Mexican Funding Agencies (CINVESTAV, CONACYT, SEP, and UASLP-FAI); the Ministry of Science and Innovation, New Zealand; the Pakistan Atomic Energy Commission; the Ministry of Science and Higher Education and the National Science Centre, Poland; the Funda\c{c}\~ao para a Ci\^encia e a Tecnologia, Portugal; JINR (Armenia, Belarus, Georgia, Ukraine, Uzbekistan); the Ministry of Education and Science of the Russian Federation, the Federal Agency of Atomic Energy of the Russian Federation, Russian Academy of Sciences, and the Russian Foundation for Basic Research; the Ministry of Science and Technological Development of Serbia; the Secretar\'{\i}a de Estado de Investigaci\'on, Desarrollo e Innovaci\'on and Programa Consolider-Ingenio 2010, Spain; the Swiss Funding Agencies (ETH Board, ETH Zurich, PSI, SNF, UniZH, Canton Zurich, and SER); the National Science Council, Taipei; the Thailand Center of Excellence in Physics, the Institute for the Promotion of Teaching Science and Technology and National Electronics and Computer Technology Center; the Scientific and Technical Research Council of Turkey, and Turkish Atomic Energy Authority; the Science and Technology Facilities Council, UK; the US Department of Energy, and the US National Science Foundation.

Individuals have received support from the Marie-Curie programme and the European Research Council (European Union); the Leventis Foundation; the A. P. Sloan Foundation; the Alexander von Humboldt Foundation; the Belgian Federal Science Policy Office; the Fonds pour la Formation \`a la Recherche dans l'Industrie et dans l'Agriculture (FRIA-Belgium); the Agentschap voor Innovatie door Wetenschap en Technologie (IWT-Belgium); the Ministry of Education, Youth and Sports (MEYS) of Czech Republic; the Council of Science and Industrial Research, India; the Compagnia di San Paolo (Torino); and the HOMING PLUS programme of Foundation for Polish Science, cofinanced from European Union, Regional Development Fund.

\bibliography{auto_generated}   % will be created by the tdr script.

\cleardoublepage \appendix\section{The CMS Collaboration \label{app:collab}}\begin{sloppypar}\hyphenpenalty=5000\widowpenalty=500\clubpenalty=5000\textbf{Yerevan Physics Institute,  Yerevan,  Armenia}\\*[0pt]
S.~Chatrchyan, V.~Khachatryan, A.M.~Sirunyan, A.~Tumasyan
\vskip\cmsinstskip
\textbf{Institut f\"{u}r Hochenergiephysik der OeAW,  Wien,  Austria}\\*[0pt]
W.~Adam, E.~Aguilo, T.~Bergauer, M.~Dragicevic, J.~Er\"{o}, C.~Fabjan\cmsAuthorMark{1}, M.~Friedl, R.~Fr\"{u}hwirth\cmsAuthorMark{1}, V.M.~Ghete, J.~Hammer, N.~H\"{o}rmann, J.~Hrubec, M.~Jeitler\cmsAuthorMark{1}, W.~Kiesenhofer, V.~Kn\"{u}nz, M.~Krammer\cmsAuthorMark{1}, I.~Kr\"{a}tschmer, D.~Liko, I.~Mikulec, M.~Pernicka$^{\textrm{\dag}}$, B.~Rahbaran, C.~Rohringer, H.~Rohringer, R.~Sch\"{o}fbeck, J.~Strauss, A.~Taurok, W.~Waltenberger, G.~Walzel, E.~Widl, C.-E.~Wulz\cmsAuthorMark{1}
% ...
\vskip\cmsinstskip
\dag:~Deceased\\
1:~~Also at Vienna University of Technology, Vienna, Austria\\
% ...

\end{sloppypar}
\end{document}